\documentclass{article}
\usepackage{physics}

\usepackage{arxiv}

\usepackage[utf8]{inputenc} 
\usepackage[T1]{fontenc}    
\usepackage{url}            
\usepackage{booktabs}       
\usepackage{amsfonts}       
\usepackage{nicefrac}       
\usepackage{microtype}      
\usepackage{lipsum}
\usepackage{graphicx}
\usepackage[colorlinks=true,
citecolor=blue,
filecolor=black,
linktoc=all,
linkcolor=magenta,
urlcolor=red]{hyperref}
\usepackage{setspace}
\graphicspath{ {./figures/} }

\title{Elementary exposition of realizing phase space structures relevant to chemical reaction dynamics}

\author{
Wenyang Lyu
\And
Shibabrat Naik
\And
Stephen Wiggins
\and
\textit{School of Mathematics, University of Bristol} \\
\textit{Fry Building, Woodland Road, Bristol BS8 1UG, United Kingdom}
}

\begin{document}
\maketitle

\begin{abstract}
    In this article, we review the analytical and numerical approaches for computing the phase space structures in two degrees-of-freedom Hamiltonian systems that arise in chemical reactions. In particular, these phase space structures are the unstable periodic orbit associated with an index-1 saddle, the periodic orbit dividing surface, and the stable and unstable invariant manifolds of the unstable periodic orbit. We review the approaches in the context of a two degrees-of-freedom Hamiltonian with a quartic potential coupled with a quadratic potential. We derive the analytical form of the phase space structures for the integrable case and visualize their geometry on the three dimensional energy surface. We then investigate the bifurcation of the dividing surface due to the changes in the parameters of the potential energy. We also review the numerical method of \emph{turning point} and present its new modification called the \emph{turning point based on configuration difference} for computing the unstable periodic orbit in two degrees-of-freedom systems. These methods are implemented in the open-source python package, UPOsHam~\cite{Lyu2020}.
\end{abstract}


\tableofcontents

\section{Introduction}

\textbf{Brief background.~}
Hamiltonian systems~\cite{Meyer2009,wiggins2003applied} appear in a wide range of areas in natural sciences and engineering, for example, chemical reactions~\cite{NHIMinchem}, celestial mechanics~\cite{koon_heteroclinic_2000,Jaffe2002,Dellnitz2005,Waalkens2005Escape}, ship dynamics~\cite{naik_geometry_2017}, and fluid dynamics~\cite{Mendoza2010}. Dynamical systems is the study of continuous (or discrete) time evolution of initial conditions governed by ordinary differential equations (ODEs), partial differential equations, or function iterations, and a Hamiltonian system is an energy conservative system. For a $N-$degrees-of-freedom system, the Hamiltonian dynamical system is described by a set of position coordinates $\mathbf{q}=(q_1, \ldots , q_n)$ and momenta $\mathbf{p}=(p_1, \ldots , p_n)$ which gives a $2N-$dimensional space called the \emph{phase space} where the initial conditions evolve in time. The time evolution of the initial conditions is governed by an autonomous or non-autonomous ordinary differential equation with respect to time called the \emph{Hamilton's equations}. Each initial condition is represented by a point in the phase space and the initial condition's past and future location in the phase space is obtained by evolving the initial condition in backward and forward time, respectively. One of the phase space structures is the equilibrium point of a Hamiltonian dynamical system, which is also called as the critical point (where the gradient vanishes) of the potential energy surface, which exists at only one value of the total energy~\cite{Wiggins2017book}. In two DOF systems, equilibrium points can be of different types and in this review, we will focus on the center $\times$ saddle (called an index-1 saddle) which corresponds to a top of the potential energy barrier and the center $\times$ center which corresponds to a potential energy well. The type of the equilibrium point is deduced by linear stability analysis of the equilibrium point. One important point to consider is that the equilibrium points have \emph{zero momentum} (this follows from the definition) and exist at one energy but their influence at non-zero momentum which corresponds to higher energies is affected by the periodic orbit associated with the equilibrium point. The periodic orbit associated with an index-1 saddle equilibrium point is classified as unstable and its generalization to high-dimensional ($N > 2$) phase space is called a Normally Hyperbolic Invariant Manifold (NHIM)~\cite{Wiggins1990,Wiggins1994}. The unstable periodic orbit (and its general form, NHIM) has been shown to be the appropriate phase space structure to construct a dividing surface that partitions the energy surface into the reactant and product regions and which is free of local recrossings~\cite{waalkens2004direct,Waalkens2005}. A phase space dividing surface leads to more accurate calculation of the reaction flux and rate constant in a chemical reaction~\cite{Ezra2009}. In this review, we illustrate the geometry of a phase space dividing surface by varying parameters of a potential energy surface. Thus, a dividing surface is the phase space structure (it is a dynamical object since momenta is taken into account) used in reaction rate calculations. In a sense this provides the local pictures of crossing the energetic barrier located by the saddle equilibrium point. However, another interest in studying chemical reactions is to test arbitrary configurations of a molecule for reactivity (potential of reaction in an environment). In this case, the global dynamics of the configurations needs to be understood and the appropriate phase space structures are the stable and unstable invariant manifolds of the unstable periodic orbit (NHIM in $N$-DOF systems) associated with the saddle equilibrium point. The stable and the unstable manifolds are codimension-1 in the energy surface and are impenetrable barriers~\cite{wiggins_impenetrable_2001} that partition trajectories with different dynamical fates.

\textbf{Hamiltonian mechanics in the context of chemical reactions.~}
In the context of chemistry, the configuration space is described by the bond lengths between atoms in a molecule and denoted by the position coordinates, $q_1,...,q_N$, in a $N-$degrees-of-freedom system. In this setup, one has to define what reaction means from a mathematical point of view. For example, a dissociation reaction involving occurs when one or more bonds break which translates to one or more position coordinates becoming unbounded when they evolve in time. These position coordinates are called the reaction coordinates. The position coordinates are also involved in describing the potential energy in a chemical reaction. Thus, the configuration space perspective is concerned with the features and shape of a potential energy function. However, the time evolution of the position coordinates, $q_i$, depends on the momenta, $p_i$, due to the kinetic energy of the bonds which arises due to the molecular motion. The combination of the potential energy and kinetic energy defines the Hamiltonian while the space in which the coordinates and momenta evolve is called the phase space~\cite{agaoglou_chemical_2019} of the Hamiltonian system. Thus, the configuration space and phase space of a molecular system are inherently related and the appropriate setting for understanding reactions is the phase space.

In this review, we illustrate the geometry of the aforementioned phase space structures by using a two DOF model Hamiltonian where all the relevant structures can be calculated analytically. It is to be noted that linear Hamiltonian systems are considered as the first step~\cite{NHIMinchem, Geomodels} in these theory since only one equilibrium point, either a saddle $\times$ center or center $\times$ center, is possible. The natural next step is to consider a model \emph{nonlinear} Hamiltonian by adding higher order terms in the potential energy function and thus allows for multiple equilibrium points. We consider a Hamiltonian (Eqn.~\eqref{defn:hamiltonian}) with a quartic term in the potential energy function, $V(q_1)$, and use this Hamiltonian to discuss the associated phase space structures along with the visualizations. In particular, we obtain analytical expressions for the NHIM which is an unstable periodic orbit associated with the index-1 saddle equilibrium point which makes the visualization feasible in the three-dimensional energy surface of the two DOF Hamiltonian (Eqn.~\eqref{defn:hamiltonian}). We demonstrate the bifurcation of the phase space DS as the parameters in the potential energy are varied and which can not happen in linear Hamiltonian systems. Next, we demonstrate numerical methods that need to be adopted when analytical expressions of the UPOs can not be obtained which is what happens in most realistic models of chemical reactions. Some of these numerical methods include differential correction and numerical continuation~\cite{naik_finding_2019b,Koon2011}, multiple shooting~\cite{farantos_pomult_1998}, turning point~\cite{Pollak1980}, and turning point based on configuration difference~\cite{Lyu2020}. We describe the turning point methods for the model Hamiltonian in this article. 

The article is outlined as follows.  First, we briefly describe the two degrees-of-freedom (DOF) model Hamiltonian in section~\ref{sect:quartic2dof} along with the basic dynamical systems analysis. Then, we obtain the phase space structures for different combinations of the signs of the parameters in the potential energy in section~\ref{subsect:phase_space}. We discuss the resulting bifurcations in section~\ref{subsect:bifurcation} and the phase space structures are visualized in section~\ref{subsect:vis}. We present the results obtained using the turning point methods in section~\ref{subsect:numeric_methods}. Then, we summarize and present our outlook in section~\ref{sect:conclusions}.

\section{Two DOF model: Uncoupled quartic Hamiltonian\label{sect:quartic2dof}}

The Hamiltonian structure is given by 
\begin{align} 
H = \underbrace{\frac{ p_1^2}{2} - \frac{\alpha q_1^2 }{2}+ \frac{\beta q_1^4}{4}}_{H_1} + \underbrace{\frac{\omega}{2}(p_2^2 + q_2 ^2)}_{H_2},\ \alpha, \ \beta \text{ are some parameters.} \label{defn:hamiltonian}
\end{align}

The Hamilton's equations of $H$ are given by:
\begin{align} 
\Dot{q_1} &= \frac{\partial H}{\partial p_1} = p_1, \label{newhamq_1} \\ 
\Dot{p_1} &= - \frac{\partial H}{\partial q_1} = \alpha q_1 - \beta q_1^3, \label{newhamp_1} \\
\Dot{q_2} &= \frac{\partial H}{\partial p_2} = \omega p_2, \label{newhamq_2} \\ 
\Dot{p_2} &= - \frac{\partial H}{\partial q_2} = -\omega q_2. \label{newhamp_2}
\end{align}

Note that ($\cdot$) stands for derivative with respect to time $t$.

Now we want to classify all equilibrium points and their stability associated to $H$. If we let
\begin{align*} 
\Dot{q_1} &= \frac{\partial H}{\partial p_1} = p_1 =0,\\ 
\Dot{p_1} &= - \frac{\partial H}{\partial q_1} = \alpha q_1 - \beta q_1^3 =0, \\
\Dot{q_2} &= \frac{\partial H}{\partial p_2} = \omega p_2=0, \\ 
\Dot{p_2} &= - \frac{\partial H}{\partial q_2} = -\omega q_2 = 0.
\end{align*}
we can see that equilibrium points are 

\begin{align}
(\bar{q_1}, \bar{p_1}, \bar{q_2}, \bar{p_2}) = (0,0,0,0), (\pm \sqrt{\alpha/\beta},0,0,0) \label{equilpts}
\end{align}

Note that $(\bar{q_1}, \bar{p_1}, \bar{q_2}, \bar{p_2}) = (0,0,0,0)$ is an equilibrium point for all $ \alpha, \beta, \omega$ and $(\bar{q_1}, \bar{p_1}, \bar{q_2}, \bar{p_2}) = (\pm \sqrt{\alpha/\beta},0,0,0)$ are equilibrium points if and only if $\alpha/\beta > 0$. However we need to consider the case $\beta = 0$ separately as the fraction $ \alpha/\beta$ does not make sense when $\beta = 0$. The eigenvalues of the Jacobian associated to $H_2$ are $\pm i \omega$ and the eigenvalues $\lambda$ of the Jacobian J associated to $H_1$ are the solutions of the characteristic equation:

\begin{align}
\textrm{det}(J - \lambda I) =  det\begin{pmatrix}
-\lambda  &        1        \\
\alpha - 3\beta q_1^2    &  -\lambda
\end{pmatrix} = \lambda^2 -(\alpha - 3\beta q_1^2) = 0
\end{align}
where
\begin{align}
J = \begin{pmatrix}
0   &   1 \\
\alpha - 3\beta q_1^2  &  0
\end{pmatrix}
\end{align}
Rearranging we have $ \lambda = \pm \sqrt{\alpha - 3\beta q_1^2} $. 

\section{Results and Discussions\label{sect:results}}

\subsection{Phase space structures\label{subsect:phase_space}}
\paragraph{1. degenerate case $ \alpha = \beta = 0$.}First, let's consider the trivial case when $ \alpha = \beta = 0$. 
Our $ H_1$ becomes $ H_1 = \frac{p_1^2}{2}$ and (\ref{newhamq_1}), (\ref{newhamp_1}) become
\begin{align*} \Dot{q_1} &= \frac{\partial H}{\partial p_1} = p_1, \\ 
\Dot{p_1} &= - \frac{\partial H}{\partial q_1} = 0.
\end{align*}
which imply that $p_1$ is a constant, i.e. phase diagram is consisted of horizontal lines in $ p_1 - q_1 $ plane.

\paragraph{2. degenerate case $ \alpha \neq 0, \beta =0$.} In this case, our Hamiltonian $H_1 $ becomes
\begin{align*} H_1 = \frac{ p_1^2}{2} - \frac{\alpha q_1^2}{2}  
\end{align*}
we only have one equilibrium point $(\bar{q_1}, \bar{p_1}) = (0,0)$ and the eigenvalues of the Jacobian are $ \lambda = \pm \sqrt{\alpha}$ . It's exactly the same to what we have discussed before.

\begin{enumerate}
    \item If $ \alpha < 0$, the eigenvalues of the Jacobian are purely imaginary, i.e. the origin is a centre. 
    \item If $ \alpha > 0$, the eigenvalues of the Jacobian are real and of opposite signs, i.e. the origin is a saddle. Therefore $(\bar{q_1}, \bar{p_1}, \bar{q_2}, \bar{p_2}) = (0,0,0,0)$ is a "saddle-centre" equilibrium point. The full energy surface $ H = H_1 + H_2$ is given by 
\begin{align*} H = \frac{ p_1^2}{2} - \frac{\alpha q_1^2}{2} + \frac{\omega}{2}(p_2^2 + q_2 ^2) = H_1 + H_2 > 0, H_1 >0, H_2 \geq 0
\end{align*}
The DS can be taken as $q_1 = 0$ and it is nonisoenergetic since
it's not consisted of trajectories in the phase space. The DS has two hemispheres which correspond to forward and backward reactions.
\begin{align*}
    (q_1, p_1, q_2, p_2 ) \in \mathbb{R}^4 : p_1 = +\sqrt{2}\sqrt{H_1+H_2- \frac{\omega}{2}(p_2^2 + q_2^2)} > 0,\ \text{forward DS}, 
\end{align*}
\begin{align*}
    (q_1, p_1, q_2, p_2 ) \in \mathbb{R}^4 : p_1 = -\sqrt{2}\sqrt{H_1+H_2- \frac{\omega}{2}(p_2^2 + q_2^2)} < 0,\ \text{backward DS}.
\end{align*}
The forward and backward DS meet at $ p_1 = 0, q_1 = 0$:
\begin{align*}
     \mathbb{S}^{1}_{\text{NHIM}}(h) = \Sigma(h) \cap \{ q_1 = 0, p_1 = 0 \} = \{ (q_1, p_1, q_2, p_2 ) \in \mathbb{R}^4 : \omega(p_2^2 + q_2^2) = 2h\} ,\ q_1 = 0,\ p_1 = 0.
\end{align*}
The NHIM is given by 
\begin{align*}  
\frac{\omega}{2}(p_2^2 + q_2^2) = H_1 + H_2 \geq 0
\end{align*}
the stable and unstable manifolds are given by 
\begin{align*}
W^u(\text{NHIM}) &= \{ (q_1, p_1, q_2, p_2) | p_1 = \sqrt{\alpha}q_1,\ \frac{\omega}{2}(p_2^2 + q_2^2) = H_2\}, \\
W^s(\text{NHIM}) &= \{ (q_1, p_1, q_2, p_2) | p_1 = -\sqrt{\alpha}q_1,\ \frac{\omega}{2}(p_2^2 + q_2^2) = H_2\}.
\end{align*}

\end{enumerate}

\paragraph{3. degenerate case $ \alpha = 0, \beta \neq 0$.} 

If $\alpha = 0, \beta \neq 0$, the Hamiltonian $ H_1$ becomes
\begin{align*} H_1 = \frac{ p_1^2}{2} +  \frac{\beta q_1^4}{4}
\end{align*}
we only have one equilibrium point, $(\bar{q_1}, \bar{p_1}) = (0,0)$ which is the origin. The eigenvalues of the Jacobian are $ \lambda = \pm \sqrt{- 3\beta q_1^2} = 0$ i.e. the equilibrium point $(\bar{q_1}, \bar{p_1}) = (0,0)$ is non hyperbolic. Therefore $(\bar{q_1}, \bar{p_1}, \bar{q_2}, \bar{p_2}) = (0,0,0,0)$ is nonhyperbolic. We can still draw the phase diagrams in $q_1-p_1$ plane depending on the signs of $\beta$. 

\begin{enumerate}
    \item If $ \beta > 0$, we can plot the potential energy function which is a "$\cup$" shape and the phase diagram is similar to the one which has centre stability. 
    \item If $ \beta< 0$, similarly we can plot the potential energy function which is a "$\cap$" shape and the phase diagram is similar to the one which has saddle stability. The invariant manifolds associated to the origin are given by 
    \begin{align*}  
    p_1^2 = -\frac{\beta}{2}q_1^4
    \end{align*}
    We can write explicitly as
    \begin{align*}
    W^c((0,0)) &= \{ (q_1, p_1) | p_1 = \sqrt{-\frac{\beta}{2}}q_1^2 \}, \\
    W^c((0,0)) &= \{ (q_1, p_1) | p_1 = -\sqrt{-\frac{\beta}{2}}q_1^2 \}.
    \end{align*}
\end{enumerate}

\paragraph{4. Nonzero $ \alpha, \beta$.} Now we consider $ \alpha, \beta \neq 0$. The Hamiltonian $H_1$ is 
\begin{align} H_1 = \frac{ p_1^2}{2} - \frac{\alpha q_1^2}{2} +  \frac{\beta q_1^4}{4}
\end{align}
Recall that the eigenvalues of the Jacobian are $ \lambda = \pm \sqrt{\alpha - 3\beta q_1^2}$. 

\begin{enumerate}
    \item If $\alpha<0, \beta >0 $, we only have 1 equilibrium point $(\bar{q_1}, \bar{p_1}) = (0,0)$ as $ \alpha/\beta$ in (\ref{equilpts}) is negative. $ \lambda = \pm \sqrt{\alpha - 3\beta q_1^2} = \pm \sqrt{\alpha}$ are purely imaginary, i.e. the origin is a centre. 
    \item If $\alpha>0, \beta <0 $, we only have 1 equilibrium point $(\bar{q_1}, \bar{p_1}) = (0,0)$ for the same reason. $ \lambda = \pm \sqrt{\alpha - 3\beta q_1^2} = \pm \sqrt{\alpha}$ are real and of opposite signs, i.e. the origin is a saddle. Therefore $(\bar{q_1}, \bar{p_1}, \bar{q_2}, \bar{p_2}) = (0,0,0,0)$ is a "saddle-centre" equilibrium point. The full energy surface $H = H_1+ H_2$ is given by 
\begin{align} H = \frac{ p_1^2}{2} - \frac{\alpha q_1^2}{2} +  \frac{\beta q_1^4}{4} + \frac{\omega}{2}(p_2^2 + q_2 ^2) = H_1 + H_2 > 0, H_1>0, H_2 \geq 0.
\end{align}
The DS can be taken as $q_1 = 0$. The DS is nonisoenergetic since
it's not consisted of trajectories in the phase space. The DS has two hemispheres which correspond to forward and backward reactions.
\begin{align}
    (q_1, p_1, q_2, p_2 ) \in \mathbb{R}^4 : p_1 = +\sqrt{2}\sqrt{H_1+H_2- \frac{\omega}{2}(p_2^2 + q_2^2)} > 0,\ \text{forward DS}, 
\end{align}
\begin{align}
    (q_1, p_1, q_2, p_2 ) \in \mathbb{R}^4 : p_1 = -\sqrt{2}\sqrt{H_1+H_2- \frac{\omega}{2}(p_2^2 + q_2^2)} < 0.\ \text{backward DS}.
\end{align}
The forward and backward DS meet at $ p_1 = 0, q_1 = 0$:
\begin{align}
     \mathbb{S}^{1}_{\text{NHIM}}(h) = \Sigma(h) \cap \{ q_1 = 0, p_1 = 0 \} = \{ (q_1, p_1, q_2, p_2 ) \in \mathbb{R}^4 : \omega(p_2^2 + q_2^2) = 2h\} ,\ q_1 = 0,\ p_1 = 0.
\end{align}
The NHIM is given by 
\begin{align}  
\frac{\omega}{2}(p_2^2 + q_2^2) = H_1 + H_2 \geq 0
\end{align}
the stable and unstable manifolds are given by 
\begin{align}
    p_1^2 = \alpha q_1^2 - \frac{\beta q_1^4}{2}
\end{align}
We can write explicitly as
\begin{align}
W^s_f(\text{NHIM}) &= \{ (q_1, p_1, q_2, p_2) | p_1 = \sqrt{\alpha q_1^2 - \frac{\beta q_1^4}{2}},\ \frac{\omega}{2}(p_2^2 + q_2^2) =H_2\},\ q_1<0,\ p_1 >0,\\
W^s_b(\text{NHIM}) &= \{ (q_1, p_1, q_2, p_2) | p_1 = -\sqrt{\alpha q_1^2 - \frac{\beta q_1^4}{2}},\ \frac{\omega}{2}(p_2^2 + q_2^2) =H_2\},\ q_1>0,\ p_1 <0,\\
W^u_f(\text{NHIM}) &= \{ (q_1, p_1, q_2, p_2) | p_1 = \sqrt{\alpha q_1^2 - \frac{\beta q_1^4}{2}},\ \frac{\omega}{2}(p_2^2 + q_2^2) =H_2\},\ q_1>0,\ p_1 >0,\\
W^u_b(\text{NHIM}) &= \{ (q_1, p_1, q_2, p_2) | p_1 = -\sqrt{\alpha q_1^2 - \frac{\beta q_1^4}{2}},\ \frac{\omega}{2}(p_2^2 + q_2^2) =H_2\},\ q_1<0,\ p_1 <0.
\end{align}
\item If $\alpha>0, \beta >0 $, we have 3 equilibrium points $(\bar{q_1}, \bar{p_1}) = (0,0), (\pm \sqrt{\alpha/\beta},0)$. The stability of these equilibrium points are determined by $ \lambda = \pm \sqrt{\alpha - 3\beta q_1^2} $. When $ q_1 = 0$, we have $ \lambda = \pm \sqrt{\alpha } $, i.e. the origin is a saddle. when $ q_1 = \pm \sqrt{\alpha/\beta}$, we have $ \lambda = \pm \sqrt{-2\alpha } $, i.e. $(\bar{q_1}, \bar{p_1}) =(\pm \sqrt{\alpha/\beta},0)$ are centres. Note that the equilibrium point $(\bar{q_1}, \bar{p_1}) = (0,0)$ exists on the energy surface $H_1 = 0$ and $ (\bar{q_1}, \bar{p_1}) =(\pm \sqrt{\alpha/\beta},0)$ exist on the energy surface $ H_1 = 0 -\frac{\alpha \alpha/\beta }{2} + \frac{\beta (\alpha/\beta)^2}{4} = -\frac{\alpha^2}{2\beta} + \frac{\alpha^2}{4\beta} = -\frac{\alpha^2}{4\beta} < 0$. The phase space structure of $H_1$ is given in Fig.~\ref{fig:phasespaceadv_alp=1_beta=1}. Note also that $(\bar{q_1}, \bar{p_1}, \bar{q_2}, \bar{p_2}) = (0,0,0,0)$ is a "saddle-centre" equilibrium point.
\begin{figure}[!h] 
    \centering \includegraphics[width=0.7\columnwidth]{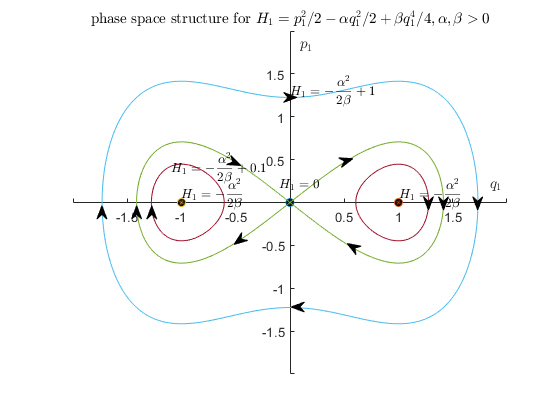}
    \caption{\label{fig:phasespaceadv_alp=1_beta=1}} Phase space structures for $\alpha, \beta = 1 $
\end{figure} 
In order for reaction to occur, we must have $H_1 > 0$. Therefore the full energy surface $H = H_1+ H_2$ is given by \begin{align} H = \frac{ p_1^2}{2} - \frac{\alpha q_1^2}{2} +  \frac{\beta q_1^4}{4} + \frac{\omega}{2}(p_2^2 + q_2 ^2) = H_1 + H_2 >0,\ H_1> 0, H_2 \geq 0.
\end{align} 
The DS can be taken as $q_1 = 0$. The DS is nonisoenergetic since
it's not consisted of trajectories in the phase space. The DS has two hemispheres which correspond to forward and backward reactions.
\begin{align}
    (q_1, p_1, q_2, p_2 ) \in \mathbb{R}^4 : p_1 = +\sqrt{2}\sqrt{H_1+H_2- \frac{\omega}{2}(p_2^2 + q_2^2)} > 0,\ \text{forward DS}, 
\end{align}
\begin{align}
    (q_1, p_1, q_2, p_2 ) \in \mathbb{R}^4 : p_1 = -\sqrt{2}\sqrt{H_1+H_2- \frac{\omega}{2}(p_2^2 + q_2^2)} < 0,\ \text{backward DS}. 
\end{align}
The forward and backward DS meet at $ p_1 = 0, q_1 = 0$:
\begin{align}
     \mathbb{S}^{1}_{\text{NHIM}}(h) = \Sigma(h) \cap \{ q_1 = 0, p_1 = 0 \} = \{ (q_1, p_1, q_2, p_2 ) \in \mathbb{R}^4 : \omega(p_2^2 + q_2^2) = 2h\} ,\ q_1 = 0,\ p_1 = 0.
\end{align}
The NHIM is given by 
\begin{align}  
\frac{\omega}{2}(p_2^2 + q_2^2) = H_1 + H_2 \geq 0
\end{align}
the stable and unstable manifolds are given by 
\begin{align}
W^s_f(\text{NHIM}) &= \{ (q_1, p_1, q_2, p_2) | p_1 = \sqrt{\alpha q_1^2 - \frac{\beta q_1^4}{2}},\ \frac{\omega}{2}(p_2^2 + q_2^2) =H_2\},\ q_1<0,\ p_1 >0,\\
W^s_b(\text{NHIM}) &= \{ (q_1, p_1, q_2, p_2) | p_1 = -\sqrt{\alpha q_1^2 - \frac{\beta q_1^4}{2}},\ \frac{\omega}{2}(p_2^2 + q_2^2) =H_2\},\ q_1>0,\ p_1 <0,\\
W^u_f(\text{NHIM}) &= \{ (q_1, p_1, q_2, p_2) | p_1 = \sqrt{\alpha q_1^2 - \frac{\beta q_1^4}{2}},\ \frac{\omega}{2}(p_2^2 + q_2^2) =H_2\},\ q_1>0,\ p_1 >0,\\
W^u_b(\text{NHIM}) &= \{ (q_1, p_1, q_2, p_2) | p_1 = -\sqrt{\alpha q_1^2 - \frac{\beta q_1^4}{2}},\ \frac{\omega}{2}(p_2^2 + q_2^2) =H_2\},\ q_1<0,\ p_1 <0.
\end{align}

\item If $\alpha<0, \beta <0 $, we have 3 equilibrium points $(\bar{q_1}, \bar{p_1}) = (0,0),\ (\pm \sqrt{\alpha/\beta},0)$. The stability of these equilibrium points are given by $ \lambda = \pm \sqrt{\alpha - 3\beta q_1^2} $. When $ q_1 = 0$, we have $ \lambda = \pm \sqrt{\alpha } $, i.e. the origin is a centre. when $ q_1 = \pm \sqrt{\alpha/\beta}$, we have $ \lambda = \pm \sqrt{-2\alpha } $, i.e. $(\bar{q_1}, \bar{p_1}) =(\pm \sqrt{\alpha/\beta},0)$ are saddles. Note that the equilibrium point $(\bar{q_1}, \bar{p_1}) = (0,0)$ exists on the energy surface $H_1 = 0$ and $ (\bar{q_1}, \bar{p_1}) =(\pm \sqrt{\alpha/\beta},0)$ exist on the energy surface $ H_1 = 0 -\frac{\alpha \alpha/\beta }{2} + \frac{\beta (\alpha/\beta)^2}{4} = -\frac{\alpha^2}{2\beta} + \frac{\alpha^2}{4\beta} = -\frac{\alpha^2}{4\beta} > 0$, i.e. the saddles exist on the energy surface which has more energy than the origin. The phase space structure of $H_1$ is given in Fig.~\ref{fig:phasespaceadv_alp=-1_beta=-1}. Note also that $(\bar{q_1}, \bar{p_1}, \bar{q_2}, \bar{p_2}) = (\pm \sqrt{\alpha/\beta},0,0,0)$ are "saddle-centre" equilibrium points.\\
\begin{figure}[!h] 
        \centering \includegraphics[width=0.7\columnwidth]{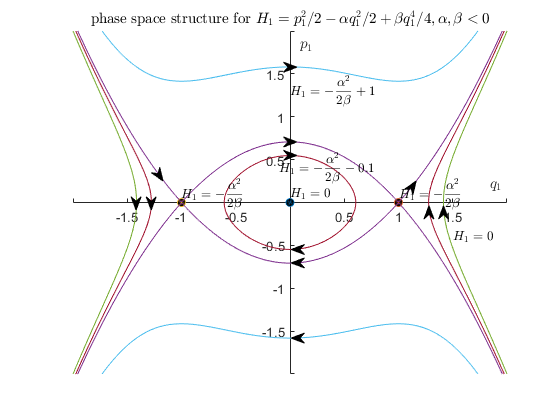}
        \caption{\label{fig:phasespaceadv_alp=-1_beta=-1}} Phase space structures for $\alpha, \beta = -1 $
\end{figure} \\
In order for reaction to occur, we must have $H_1 > -\frac{\alpha^2}{4\beta} > 0$. Therefore the full energy surface $H = H_1+ H_2$ is given by \begin{align} H = \frac{ p_1^2}{2} - \frac{\alpha q_1^2}{2} +  \frac{\beta q_1^4}{4} + \frac{\omega}{2}(p_2^2 + q_2 ^2) = H_1 + H_2 > -\frac{\alpha^2}{4\beta},\ H_1> -\frac{\alpha^2}{4\beta} > 0, H_2 \geq 0.
\end{align}
The DS can be taken as $q_1 = \sqrt{\alpha/\beta}$ or $q_1 = -\sqrt{\alpha/\beta}$. The DS is nonisoenergetic since
it's not consisted of trajectories in the phase space. Each DS has two hemispheres which correspond to forward and backward reactions.
\begin{align}
    (q_1, p_1, q_2, p_2 ) \in \mathbb{R}^4 : p_1 = +\sqrt{2}\sqrt{H_1+H_2- \frac{\omega}{2}(p_2^2 + q_2^2) +\frac{\alpha^2}{4\beta}} > 0,\ q_1 = \sqrt{\alpha/\beta}, \\ \text{forward DS}(q_1 = \sqrt{\alpha/\beta}), 
\end{align}
\begin{align}
    (q_1, p_1, q_2, p_2 ) \in \mathbb{R}^4 : p_1 = -\sqrt{2}\sqrt{H_1+H_2- \frac{\omega}{2}(p_2^2 + q_2^2)+\frac{\alpha^2}{4\beta}} < 0,\ q_1 = \sqrt{\alpha/\beta},\\ \text{backward DS}(q_1 = \sqrt{\alpha/\beta}), 
\end{align}
\begin{align}
    (q_1, p_1, q_2, p_2 ) \in \mathbb{R}^4 : p_1 = +\sqrt{2}\sqrt{H_1+H_2- \frac{\omega}{2}(p_2^2 + q_2^2) +\frac{\alpha^2}{4\beta}} > 0,\ q_1 = -\sqrt{\alpha/\beta}, \\ \text{forward DS}(q_1 = -\sqrt{\alpha/\beta}), 
\end{align}
\begin{align}
    (q_1, p_1, q_2, p_2 ) \in \mathbb{R}^4 : p_1 = -\sqrt{2}\sqrt{H_1+H_2- \frac{\omega}{2}(p_2^2 + q_2^2)+\frac{\alpha^2}{4\beta}} < 0,\ q_1 = -\sqrt{\alpha/\beta},\\ \text{backward DS}(q_1 = -\sqrt{\alpha/\beta}). 
\end{align}
The forward and backward DS meet at $ p_1 = 0, q_1 = \sqrt{\alpha/\beta}$ or $p_1 = 0, q_1 = -\sqrt{\alpha/\beta}$:
\begin{align}
     \mathbb{S}^{1}_{\text{NHIM}}(h) = \Sigma(h) \cap \{ q_1 = \pm \sqrt{\alpha/\beta}, p_1 = 0 \} = \{ (q_1, p_1, q_2, p_2 ) \in \mathbb{R}^4 : \omega(p_2^2 + q_2^2) = 2h +\frac{\alpha^2}{2\beta} \} ,\\ q_1 = \pm \sqrt{\alpha/\beta}, p_1 = 0.
\end{align}
The NHIM is given by 
\begin{align}  
\frac{\omega}{2}(p_2^2 + q_2^2) = H_1 + H_2 \geq -\frac{\alpha^2}{4\beta}
\end{align}
the invariant manifolds of the NHIMs are given by
\begin{align}  
 -\frac{\alpha^2}{4\beta} = \frac{ p_1^2}{2} - \frac{\alpha q_1^2}{2} +  \frac{\beta q_1^4}{4}
\end{align}
the stable and unstable manifolds are given by 
\begin{align} 
W^s(\text{NHIM with }q_1= \sqrt{\alpha/\beta},q_2= 0) &= \{ (q_1, p_1, q_2, p_2) | p_1 = \sqrt{\alpha q_1^2 - \frac{\beta q_1^4}{2}-\frac{\alpha^2}{2\beta}},\ \frac{\omega}{2}(p_2^2 + q_2^2) =H_2\},\ q_1<0,\\
W^u(\text{NHIM with }q_1= \sqrt{\alpha/\beta},q_2= 0) &= \{ (q_1, p_1, q_2, p_2) | p_1 = -\sqrt{\alpha q_1^2 - \frac{\beta q_1^4}{2}-\frac{\alpha^2}{2\beta}},\ \frac{\omega}{2}(p_2^2 + q_2^2) =H_2\},\ q_1<0,\\
W^s(\text{NHIM with }q_1= -\sqrt{\alpha/\beta},q_2= 0) &= \{ (q_1, p_1, q_2, p_2) | p_1 = -\sqrt{\alpha q_1^2 - \frac{\beta q_1^4}{2}-\frac{\alpha^2}{2\beta}},\ \frac{\omega}{2}(p_2^2 + q_2^2) =H_2\},\ q_1>0,\\
W^u(\text{NHIM with }q_1= -\sqrt{\alpha/\beta},q_2= 0) &= \{ (q_1, p_1, q_2, p_2) | p_1 = \sqrt{\alpha q_1^2 - \frac{\beta q_1^4}{2}-\frac{\alpha^2}{2\beta}},\ \frac{\omega}{2}(p_2^2 + q_2^2) =H_2\},\ q_1>0.\\
\end{align}
\end{enumerate}

\subsection{Bifurcation analysis\label{subsect:bifurcation}}
\begin{figure}[!ht]
	\centering
	\includegraphics[width=0.49\textwidth]{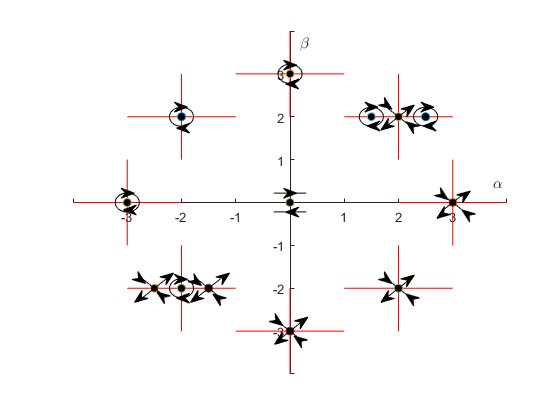}
	\includegraphics[width=0.49\textwidth]{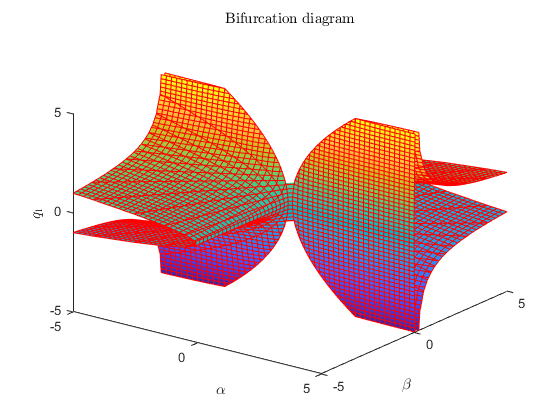}
	\caption{\textbf{Bifurcation diagram.} LHS is in $\alpha-\beta$ plane and RHS is in $\alpha-\beta-q_1$ plane. }\label{fig: bifurcation diagram}
\end{figure}

We have discussed the number and stability of the equilibrium points for different signs of $ \alpha, \beta$ and therefore we can plot the Bifurcation diagram of the Hamiltonian $H_1$ in $ \beta- \alpha$ plane in Fig.~\ref{fig: bifurcation diagram}.

When $\alpha< 0$, $\beta > 0$, we have 1 centre equilibrium point, the origin. If we fix $\alpha$ and decrease the value of $\beta$, we still have 1 equilibrium point until $\beta = 0$. As $ \beta$ becomes negative, the origin remains a centre and we have 2 new saddle equilibrium points at $ (q_1, p_1) = (\pm \sqrt{\alpha/\beta},0)$. In this case, our parameters $\alpha <0$ and $ \beta < 0$. As the number of equilibrium points change from 1 to 3, we have Hamiltonian Pitchfork Bifurcation~\cite{Wiggins2017book}. 

Now we fix $\beta >0$ and increase the value $\alpha$, we have 3 equilibrium points until $\alpha = 0$. As $\alpha$ evolves from negative to positive, $ (q_1, p_1) = (\pm \sqrt{\alpha/\beta},0)$ disappear and the origin is non hyperbolic when $\alpha =0$ and becomes a hyperbolic saddle point when $\alpha >0$. When $\beta < 0$, $ \alpha > 0$, we only have 1 equilibrium point and therefore we have Hamiltonian Pitchfork Bifurcation. 

If we fix $\alpha > 0$ and increase $\beta$ from negative to positive, we have 3 equilibrium points. The origin remains a saddle and two centres occur at $ (q_1, p_1) = (\pm \sqrt{\alpha/\beta},0)$. In this case, $\alpha>0, \beta > 0$, we still have Hamiltonian Pitchfork Bifurcation. 

Finally, if we fix $\beta >0$ and let $\alpha$ decrease from positive to negative. We have 3 equilibrium points until $\alpha = 0$. The origin loses its hyperbolicity at $\alpha = 0$. When $\alpha < 0$, $\beta > 0$, we have 1 centre equilibrium point again. Hamiltonian Pitchfork Bifurcation occurs for the same reason as above. 

\subsection{Visualisation of the phase space structures\label{subsect:vis}}

In this section we use Matlab to simulate various phase structures associated to the Hamiltonian $H= H_1 + H_2$ discussed before. We are only interested in the case $ \alpha >0, \beta > 0$ and $ \alpha < 0, \beta < 0$ and details of other cases can be found in ~\cite{Geomodels,NHIMinchem}. The method we use to visualise the phase space structures is model I in ~\cite{Geomodels} where $\Sigma_{\pm}(h)$ is the energy surface with total energy $H=H_1+H_2=h$. The sign of $p_1 = + \sqrt{2h - 2H_2-(-\alpha q_1^2+\beta^4/2)}$ is postive on $\Sigma_{+}(h)$ and the sign of $p_1 = - \sqrt{2h - 2H_2-(-\alpha q_1^2+\beta^4/2)}$ is negative on $\Sigma_{-}(h)$. Given an initial condition $(q_1,p_1,q_2,p_2) = (q_1^0,p_1^0,q_2^0,p_2^0)$, the trajectory evolves on an invariant manifold:
\begin{align}
    \Lambda_{I,J}(h) = \{ (q_1, p_1, q_2, p_2 ) \in \mathbb{R}^4 : H_1=\frac{ (p_1^0)^2}{2} - \frac{\alpha (q_1^0)^2 }{2}+\frac{\beta (q_1^0)^4}{4}=I,H_2=  \frac{\omega}{2}((p_2^0)^2 + (q_2^0) ^2), H = I+J = h\}
\end{align}
\subsubsection{Case I: \texorpdfstring{$\alpha>0, \beta>0$}{Lg}}

\begin{figure}[h!]
\centering
\begin{subfigure}[b]{0.49\linewidth}
\includegraphics[width=\linewidth]{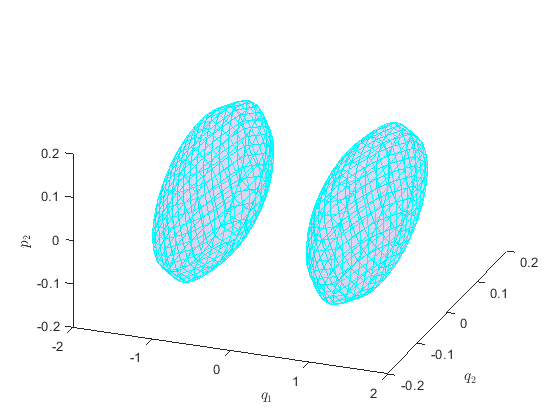}
\caption{Model I: the energy surface $\Sigma_{+}(h)$ for $h = -0.1$. }
\end{subfigure}
\begin{subfigure}[b]{0.49\linewidth}
\includegraphics[width=\linewidth]{Advanced_alpha=1_beta=1_omega=10_h=minus0dot1_EnergySurf.png}
\caption{Model I: the energy surface $\Sigma_{-}(h)$ for $h = -0.1$. }
\end{subfigure}
\begin{subfigure}[b]{0.49\linewidth}
\includegraphics[width=\linewidth]{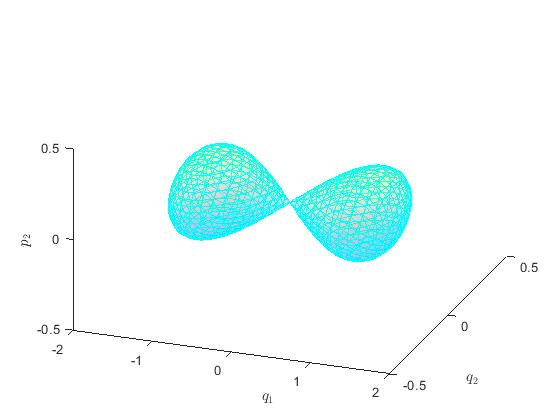}
\caption{Model I: the energy surface $\Sigma_{+}(h)$ for $h = 0$. }
\end{subfigure}
\begin{subfigure}[b]{0.49\linewidth}
\includegraphics[width=\linewidth]{Advanced_alpha=1_beta=1_omega=10_h=0_EnergySurf.png}
\caption{Model I: the energy surface $\Sigma_{-}(h)$ for $h = 0$. }
\end{subfigure}
\begin{subfigure}[b]{0.49\linewidth}
\includegraphics[width=\linewidth]{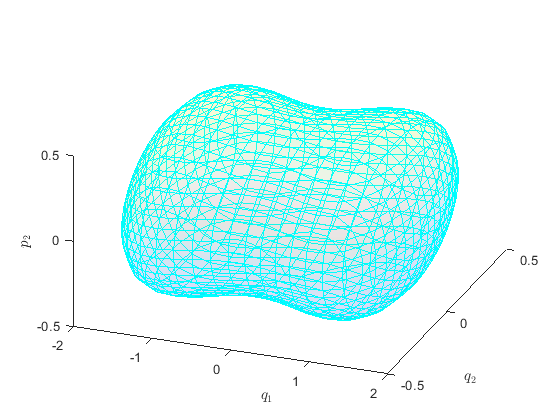}
\caption{Model I: the energy surface $\Sigma_{+}(h)$ for $h = 1$. }
\end{subfigure}
\begin{subfigure}[b]{0.49\linewidth}
\includegraphics[width=\linewidth]{Advanced_alpha=1_beta=1_omega=10_h=1_EnergySurf.png}
\caption{Model I: the energy surface $\Sigma_{-}(h)$ for $h = 1$. }
\end{subfigure}
\caption{Model I: the energy surface $\Sigma_{+,-}(h)$ for $h = -1, 0$ and$ 1$. $\alpha, \beta$, $\omega$ are chosen to be $1, 1, 10$. }
\label{fig:advancedenergy surface_alpha=1_beta=1}
\end{figure}

\begin{figure}[h!]
\centering
\begin{subfigure}[b]{0.42\linewidth}
\includegraphics[width=\linewidth]{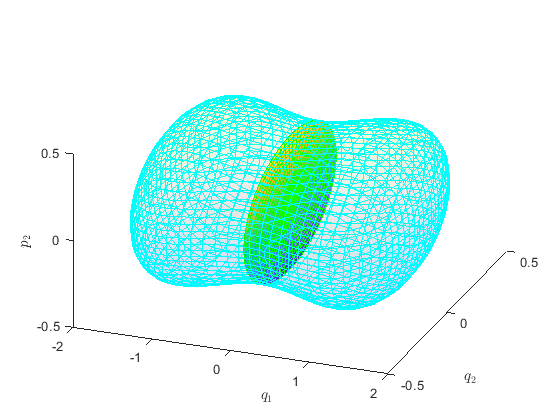}
\caption{Model I: northern hemisphere $ B^2_{ds,f}(h)$ of DS(green) which is contained in the energy surface $\Sigma_{+}(h)$(cyan). }
\end{subfigure}
\begin{subfigure}[b]{0.42\linewidth}
\includegraphics[width=\linewidth]{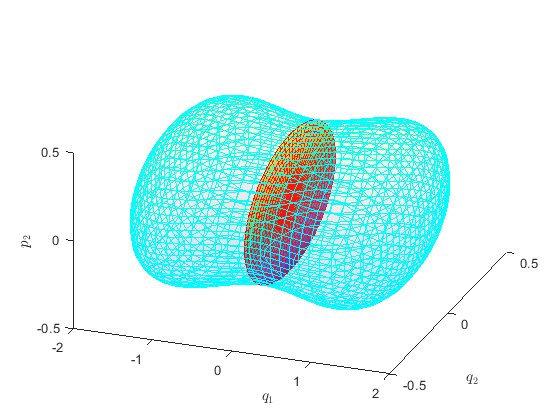}
\caption{Model I: southern hemisphere $ B^2_{ds,b}(h)$ of DS(red) which is contained in the energy surface $\Sigma_{-}(h)$(cyan). }
\end{subfigure}
\begin{subfigure}[b]{0.42\linewidth}
\includegraphics[width=\linewidth]{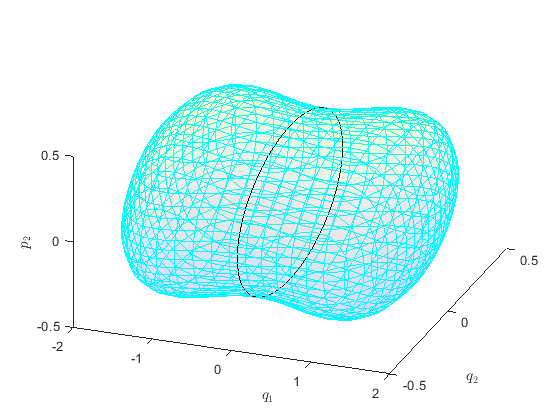}
\caption{Model I: NHIM(black) which is contained in the energy surface $\Sigma_{+}(h)$(cyan). }
\end{subfigure}
\begin{subfigure}[b]{0.42\linewidth}
\includegraphics[width=\linewidth]{Advanced_alpha=1_beta=1_omega=10_h=1_EnergySurf_NHIM.png}
\caption{Model I: NHIM(black) which is contained in the energy surface $\Sigma_{-}(h)$(cyan). }
\end{subfigure}
\caption{Model I: the dividing surface(DS) and the NHIM. Here the energy surface $\Sigma_{+,-}(h)$ for h=1. $\alpha, \beta$, $\omega$ are chosen to be $1, 10$. }
\label{fig:advanced DS_NHIM_alpha=1_beta=1}
\end{figure}

First, we present the structure of the energy surface with 3 different $h$ in Fig.~\ref{fig:advancedenergy surface_alpha=1_beta=1}. We choose the region $q_1<0$ to represent the reactants region and $q_1> 0$ to represent the products region. We can see from the graphs that in order for a reaction to occur, we must have $h > 0$. This is because when $ h \leq 0$, no trajectories cross the DS. Note that the DS is the intersection of the energy surface with the plane \{ $q_1 = 0$ \} in Fig.~\ref{fig:advanced DS_NHIM_alpha=1_beta=1}a,b and the NHIM forms the equator of the DS in Fig.~\ref{fig:advanced DS_NHIM_alpha=1_beta=1}c,d. Note also that the NHIM exists on the boundary of the energy surface.
Due to the nature of our Hamiltonian $H_1$, our energy surface only exist for $H = H_1 + H_2 \geq -\frac{\alpha^2}{4\beta}$. When $H = -\frac{\alpha^2}{4\beta}$, we have $H_1 =-\frac{\alpha^2}{4\beta}, H_2 = 0 $ which correspond to two points at $(q_1,p_1,q_2,p_2) = (\pm \sqrt{\alpha/\beta},0,0,0)$.

\begin{figure}[h!]
  \centering
  \begin{subfigure}[b]{0.42\linewidth}
    \includegraphics[width=\linewidth]{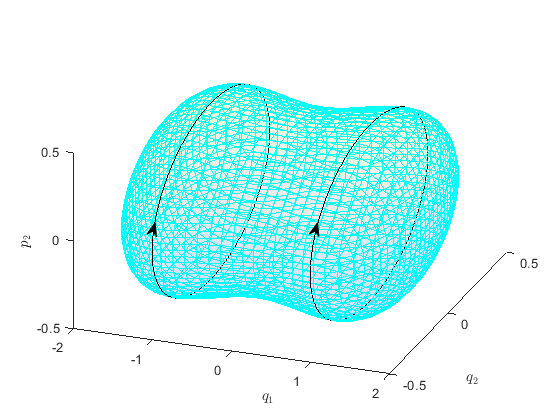}
    \caption{Model I: nonreactive trajectory(black) that is contained in invariant manifolds(yellow) $\Lambda_{I,J}(h)$ for $I= -0.25$, $J= 1.25$. Trajectory(left) is contained in the reactants component and trajectory(right) is contained in the products component. Invariant manifold $\Lambda_{I,J}(h)$ are contained in $\Sigma_{+}(h)$(cyan). }
  \end{subfigure}
  \begin{subfigure}[b]{0.42\linewidth}
    \includegraphics[width=\linewidth]{Advanced_alpha=1_beta=1_omega=10_I=minus0dot25_J=1dot25_LagCylin_Traj.png}
    \caption{Model I: nonreactive trajectory(black) that is contained in a invariant manifold(yellow) $\Lambda_{I,J}(h)$ for $I= -0.25$, $J= 1.25$. Trajectory(left) is contained in the reactants component and trajectory(right) is contained in the products component. Invariant manifolds $\Lambda_{I,J}(h)$ are contained in $\Sigma_{-}(h)$(cyan).}
  \end{subfigure}
  \caption{Model I: nonreactive trajectories(black). Here $I= -0.25$, $J= 1.25$. $\alpha, \beta$, $\omega$ are chosen to be $1, 1, 10$. Direction of the trajectories are shown by black arrows.}
  \label{fig: Advanced_alpha=1_beta=1_omega=10_I=minus0dot25_J=1dot25_LagCylin_Traj}
\end{figure}

In Fig.~\ref{fig: Advanced_alpha=1_beta=1_omega=10_I=minus0dot25_J=1dot25_LagCylin_Traj}a, we present 2 nonreactive trajectories that are contained in $\Lambda_{I,J}(h)$ for $ I = -\frac{\alpha^2}{4\beta} < 0$. The invariant manifold consists of two separate parts and is contained in $\Sigma_{+}(h)$(cyan). The left part is in the reactants component and the right part is in the products component. The dynamics in $q_1 - p_1$ plane correspond to the two equilibrium points on $ (q_1, p_1) = (\pm \sqrt{\alpha/\beta},0)$. Hence the invariant manifold consists of 2 periodic orbits in $ q_1-q_2-p_2$ plane. Any trajectory starts on each periodic orbits will stay on the periodic orbit for all time. Note that the invariant manifold in this case exists on the boundary of the energy surface. 

Similarly, we present 2 nonreactive trajectories that are contained in $\Lambda_{I,J}(h)$ for $ I = -\frac{\alpha^2}{4\beta} < 0$ in Fig.~\ref{fig: Advanced_alpha=1_beta=1_omega=10_I=minus0dot25_J=1dot25_LagCylin_Traj}b. The invariant manifold consists of two separate parts and is contained in $\Sigma_{-}(h)$(cyan). The left part is in the reactants component and the right part is in the products component. The dynamics in $q_1 - p_1$ plane correspond to the two equilibrium points on $ (q_1, p_1) = (\pm \sqrt{\alpha/\beta},0)$. Hence the invariant manifold consists of 2 periodic orbits in $ q_1-q_2-p_2$ plane. Any trajectory starts on each periodic orbits will stay on the periodic orbit for all time. Note that the invariant manifold in this case exists on the boundary of the energy surface.

\begin{figure}[h!]
  \centering
  \begin{subfigure}[b]{0.42\linewidth}
    \includegraphics[width=\linewidth]{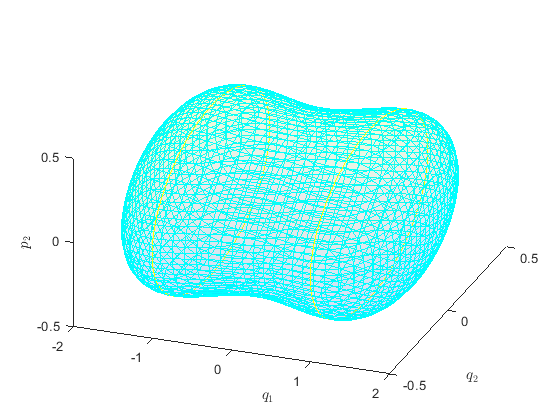}
    \caption{Model I: Invariant manifolds $\Lambda_{I,J}(h)$ are contained in $\Sigma_{+}(h)$(cyan). Invariant manifold (left) is contained in the reactants component and invariant manifold(right) is contained in the products component.}
  \end{subfigure}
  \begin{subfigure}[b]{0.42\linewidth}
    \includegraphics[width=\linewidth]{Advanced_alpha=1_beta=1_omega=10_I=minus0dot25_J=1dot25_LagCylin.png}
    \caption{Model I: Invariant manifolds $\Lambda_{I,J}(h)$ are contained in $\Sigma_{-}(h)$(cyan). Invariant manifold (left) is contained in the reactants component and invariant manifold(right) is contained in the products component.}
  \end{subfigure}
  \caption{Model I: Invariant manifolds(yellow). Here $I= -0.25$, $J= 1.25$. $\alpha, \beta$, $\omega$ are chosen to be $1, 1, 10$.}
  \label{fig: Advanced_alpha=1_beta=1_omega=10_I=minus0dot25_J=1dot25_LagCylin}
\end{figure}

For clarity, we present the invariant manifolds separately in Fig.~\ref{fig: Advanced_alpha=1_beta=1_omega=10_I=minus0dot25_J=1dot25_LagCylin}. 

\begin{figure}[h!]
  \centering
  \begin{subfigure}[b]{0.42\linewidth}
    \includegraphics[width=\linewidth]{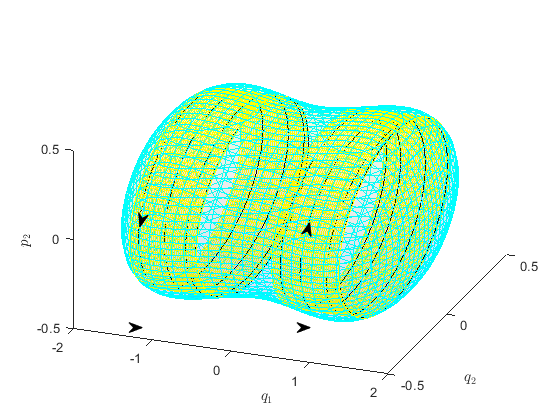}
    \caption{Model I: nonreactive trajectory(black) that is contained in invariant manifolds(yellow) $\Lambda_{I,J}(h)$ for $I= -0.1$, $J= 1.1$. Trajectory(left) is contained in the reactants component and trajectory(right) is contained in the products component. Invariant manifolds $\Lambda_{I,J}(h)$ are contained in $\Sigma_{+}(h)$(cyan). The arrows outside the Energy surface represent the trajectories oscillate from left to right. }
  \end{subfigure}
  \begin{subfigure}[b]{0.42\linewidth}
    \includegraphics[width=\linewidth]{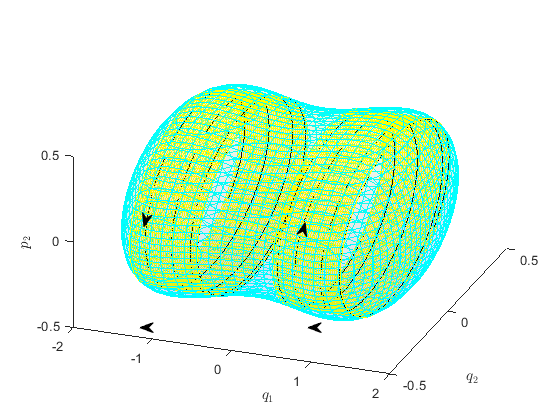}
    \caption{Model I: nonreactive trajectory(black) that is contained in invariant manifolds(yellow) $\Lambda_{I,J}(h)$ for $I= -0.1$, $J= 1.1$. Trajectory(left) is contained in the reactants component and trajectory(right) is contained in the products component. Invariant manifolds $\Lambda_{I,J}(h)$ are contained in $\Sigma_{-}(h)$(cyan). The arrows outside the Energy surface represent the trajectories oscillate from right to left. }
  \end{subfigure}
  \caption{Model I: nonreactive trajectories(black). Here $I= -0.1$, $J= 1.1$. $\alpha, \beta$, $\omega$ are chosen to be 1, 1, 10. Direction of the trajectories are shown by black arrows. The arrows outside the energy surface represent an overall direction of the trajectories.}
  \label{fig: Advanced_alpha=1_beta=1_omega=10_I=minus0dot1_J=1dot1_LagCylin_Traj_forward}
\end{figure}

In Fig.~\ref{fig: Advanced_alpha=1_beta=1_omega=10_I=minus0dot1_J=1dot1_LagCylin_Traj_forward}a, we present 2 nonreactive trajectories that are contained in $\Lambda_{I,J}(h)$ for $ -\frac{\alpha^2}{4\beta} < I < 0$. The invariant manifold consists of two separate parts and is contained in $\Sigma_{+}(h)$(cyan). The left part is in the reactants component and the right part is in the products component. 

In Fig.~\ref{fig: Advanced_alpha=1_beta=1_omega=10_I=minus0dot1_J=1dot1_LagCylin_Traj_forward}b, we present 2 nonreactive trajectories that are contained in $\Lambda_{I,J}(h)$ for $ -\frac{\alpha^2}{4\beta} < I < 0$. The invariant manifold consists of two separate parts and is contained in $\Sigma_{-}(h)$(cyan). The left part is in the reactants component and the right part is in the products component. 

The direction of the trajectories are determined in the following: 

Suppose that a trajectory starts on the invariant manifold in the reactants component(left part of the invariant manifold in Fig.~\ref{fig: Advanced_alpha=1_beta=1_omega=10_I=minus0dot1_J=1dot1_LagCylin_Traj_forward}a, it oscillates to the right until it reaches the boundary of the energy surface $\Sigma_{+}(h)$(right end of the invariant manifold in the left), where $p_1=0$ and it jumps to $\Sigma_{-}(h)$(right end of the invariant manifold in the left in Fig.~\ref{fig: Advanced_alpha=1_beta=1_omega=10_I=minus0dot1_J=1dot1_LagCylin_Traj_forward}b. It then oscillates to the left until it reaches the boundary of the energy surface $\Sigma_{-}(h)$(left end of the invariant manifold in the left), where $p_1=0$ and it jumps back to $\Sigma_{+}(h)$(left end of the invariant manifold in the left in Fig.~\ref{fig: Advanced_alpha=1_beta=1_omega=10_I=minus0dot1_J=1dot1_LagCylin_Traj_forward}a). It then oscillates to the right and we have the same motion again.

Similarly, suppose that a trajectory starts on the invariant manifold in the products component(right part of the invariant manifold in Fig.~\ref{fig: Advanced_alpha=1_beta=1_omega=10_I=minus0dot1_J=1dot1_LagCylin_Traj_forward}a, it oscillates to the right until it reaches the boundary of the energy surface $\Sigma_{+}(h)$(right end of the invariant manifold in the right), where $p_1=0$ and it jumps to $\Sigma_{-}(h)$(right end of the invariant manifold in the right in Fig.~\ref{fig: Advanced_alpha=1_beta=1_omega=10_I=minus0dot1_J=1dot1_LagCylin_Traj_forward}b. It then oscillates to the left until it reaches the boundary of the energy surface $\Sigma_{-}(h)$(left end of the invariant manifold in the right), where $p_1=0$ and it jumps back to $\Sigma_{+}(h)$(left end of the invariant manifold in the right in Fig.~\ref{fig: Advanced_alpha=1_beta=1_omega=10_I=minus0dot1_J=1dot1_LagCylin_Traj_forward}a. It then oscillates to the right and we have the same motion again.

No reaction occurs in this case as the trajectories oscillate back before they reach the NHIM. Any trajectories start in the reactants component oscillate in the reactants component for all time and any trajectories start in the products component oscillate in the products component for all time. The overall directions of the trajectories are shown by the arrows outside the energy surface.

\begin{figure}[h!]
  \centering
  \begin{subfigure}[b]{0.42\linewidth}
    \includegraphics[width=\linewidth]{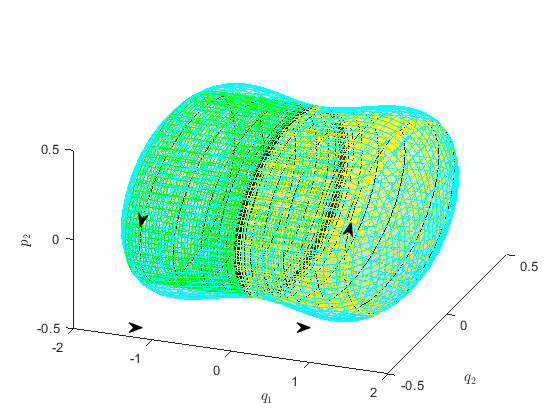}
    \caption{Model I: forward branches $W^{s}_{f}$(green) and $W^{u}_{f}$(yellow) is contained in $\Sigma_{+}(h)$(cyan). }
  \end{subfigure}
  \begin{subfigure}[b]{0.42\linewidth}
    \includegraphics[width=\linewidth]{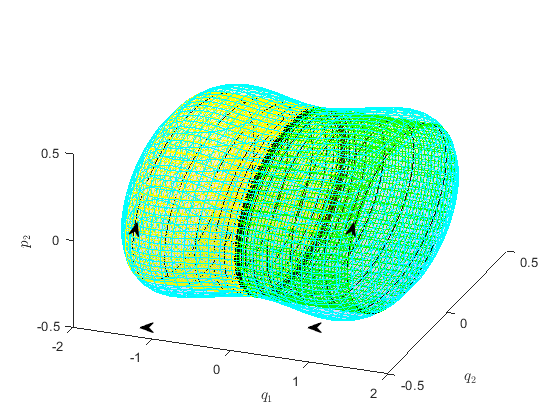}
    \caption{Model I: backward branches $W^{s}_{b}$(green) and $W^{u}_{b}$(yellow) that is contained in $\Sigma_{-}(h)$(cyan). }
  \end{subfigure}
  \caption{Model I: stable and unstable manifolds of the NHIM. Here $I = 0$, $J= 1$. $\alpha, \beta$, $\omega$ are chosen to be 1, 1, 10. Direction of the trajectories are shown by black arrows. The arrows outside the energy surface represent an overall direction of the trajectories.}
  \label{Advanced_alpha=1_beta=1_omega=10_I=0_J=1_LagCylin_Traj_Sta_Unsta}
\end{figure}

Fig.~\ref{Advanced_alpha=1_beta=1_omega=10_I=0_J=1_LagCylin_Traj_Sta_Unsta} shows the stable and unstable manifolds of the NHIM. In our case, the NHIM is a periodic orbit which forms the equator of the DS and runs around the boundaries of $\Sigma_{-}(h)$ and $\Sigma_{+}(h)$. In Fig.~\ref{Advanced_alpha=1_beta=1_omega=10_I=0_J=1_LagCylin_Traj_Sta_Unsta}a, the trajectory which starts on the $W^{s}_{f}$(reactants region) approaches the NHIM as $ t \rightarrow \infty$. The trajectory which starts on the $W^{u}_{f}$(products region) approaches the NHIM as $ t \rightarrow -\infty$. 
However, it's interesting to see that if a trajectory starts on the $W^{u}_{f}$ in Fig.~\ref{Advanced_alpha=1_beta=1_omega=10_I=0_J=1_LagCylin_Traj_Sta_Unsta}a, it oscillates to the right until it reaches the boundary of the energy surface $\Sigma_{+}(h)$, where $ p_1 = 0$. It then jumps to $W^{s}_{b}$(products region) in Fig.~\ref{Advanced_alpha=1_beta=1_omega=10_I=0_J=1_LagCylin_Traj_Sta_Unsta}b and oscillates to the left towards the NHIM. \\

In Fig.~\ref{Advanced_alpha=1_beta=1_omega=10_I=0_J=1_LagCylin_Traj_Sta_Unsta}b, the trajectory which starts on the $W^{s}_{b}$(products region) approaches the NHIM as $ t \rightarrow \infty$. The trajectory which starts on the $W^{u}_{b}$(reactants region) approaches the NHIM as $ t \rightarrow -\infty$.
It's also interesting to see that if a trajectory starts on the $W^{u}_{b}$(reactants region) in Fig.~\ref{Advanced_alpha=1_beta=1_omega=10_I=0_J=1_LagCylin_Traj_Sta_Unsta}b, it oscillates to the left until it reaches the boundary of the energy surface $\Sigma_{-}(h)$, where $ p_1 = 0$. It then jumps to $W^{s}_{f}$(reactants region) in Fig.~\ref{Advanced_alpha=1_beta=1_omega=10_I=0_J=1_LagCylin_Traj_Sta_Unsta}a and oscillates to the right towards the NHIM.

\begin{figure}[h!]
  \centering
  \begin{subfigure}[b]{0.42\linewidth}
    \includegraphics[width=\linewidth]{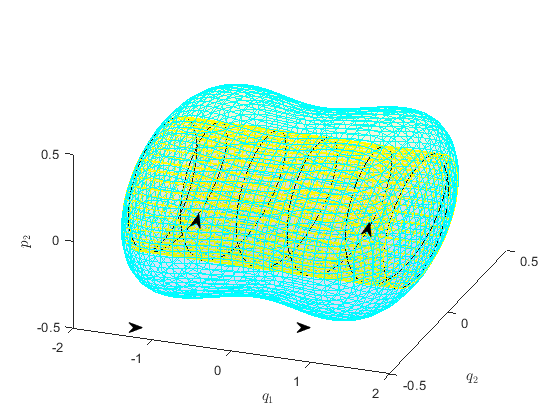}
    \caption{Model I: forward reactive trajectory(black) that is contained in a invariant manifold(yellow) $\Lambda_{I,J}(h)$ for $I = 0.5$, $J= 0.5$. Invariant manifold $\Lambda_{I,J}(h)$ is contained in $\Sigma_{+}(h)$(cyan). }
  \end{subfigure}
  \begin{subfigure}[b]{0.42\linewidth}
    \includegraphics[width=\linewidth]{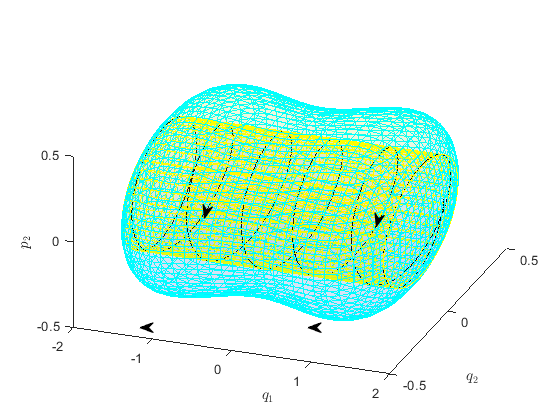}
    \caption{Model I: Backward reactive trajectory(black) that is contained in a invariant manifold(yellow) $\Lambda_{I,J}(h)$ for $I = 0.5$, $J= 0.5$. Invariant manifold $\Lambda_{I,J}(h)$ is contained in $\Sigma_{-}(h)$(cyan). }
  \end{subfigure}
  \caption{Model I: reactive trajectories contained in the invariant manifolds. Here $I = 0.5$, $J= 0.5$. $\alpha, \beta$, $\omega$ are chosen to be 1, 1, 10. Direction of the trajectories are shown by black arrows.}
  \label{Advanced_alpha=1_beta=1_omega=10_I=0dot5_J=0dot5_LagCylin_Traj}
\end{figure}

In Fig.~\ref{Advanced_alpha=1_beta=1_omega=10_I=0dot5_J=0dot5_LagCylin_Traj} we present a forward in Fig.~\ref{Advanced_alpha=1_beta=1_omega=10_I=0dot5_J=0dot5_LagCylin_Traj}a and a backward in Fig.~\ref{Advanced_alpha=1_beta=1_omega=10_I=0dot5_J=0dot5_LagCylin_Traj}b reactive trajectory that are contained in invariant manifolds. In Fig.~\ref{Advanced_alpha=1_beta=1_omega=10_I=0dot5_J=0dot5_LagCylin_Traj}a, the trajectory comes from reactants region($q_1 < 0$), approaches the DS($q_1= 0$) and then enters products region($q_1 >0$). It's therefore characterized as a forward reactive trajectory. The trajectory continues to oscillate until it reaches the boundary of the energy surface $\Sigma_{+}(h)$, where $ p_1 = 0$. It then jumps to the products region in Fig.~\ref{Advanced_alpha=1_beta=1_omega=10_I=0dot5_J=0dot5_LagCylin_Traj}b($q_1 > 0$), approaches the DS($q_1= 0$) and then enters reactants region($q_1 <0$). It's therefore characterized as a backward reactive trajectory. It goes to the left until it reaches the boundary of the energy surface $\Sigma_{-}(h)$, where $ p_1 = 0$. It then jumps to the reactants region in Fig.~\ref{Advanced_alpha=1_beta=1_omega=10_I=0dot5_J=0dot5_LagCylin_Traj}a($q_1 < 0$) and we have the same motion again. We can think of this as a reversible reaction that keeps changing from reactants to products, and from products to reactants. 

\subsubsection{Case II: \texorpdfstring{$\alpha<0, \beta<0$}{Lg}}
\begin{figure}[h!]
  \centering
  \begin{subfigure}[b]{0.42\linewidth}
    \includegraphics[width=\linewidth]{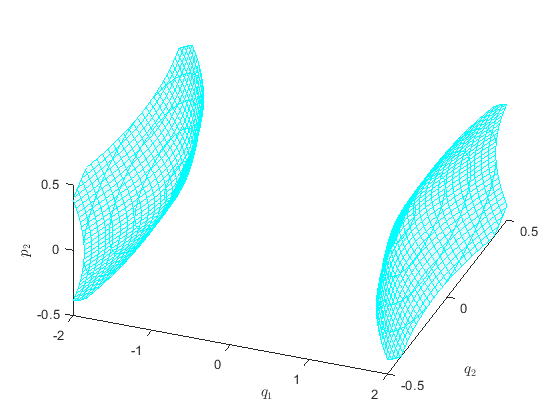}
    \caption{Model I: the energy surface $\Sigma_{+}(h)$ for h = 0. }
  \end{subfigure}
  \begin{subfigure}[b]{0.42\linewidth}
    \includegraphics[width=\linewidth]{Advanced_alpha=minus1_beta=minus1_omega=10_h=0_EnergySurf.png}
    \caption{Model I: the energy surface $\Sigma_{-}(h)$ for h = 0. }
  \end{subfigure}
  \begin{subfigure}[b]{0.42\linewidth}
    \includegraphics[width=\linewidth]{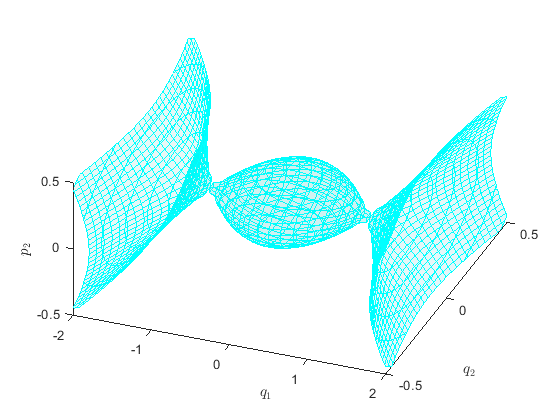}
    \caption{Model I: the energy surface $\Sigma_{+}(h)$ for h = 0.25. }
  \end{subfigure}
  \begin{subfigure}[b]{0.42\linewidth}
    \includegraphics[width=\linewidth]{Advanced_alpha=minus1_beta=minus1_omega=10_h=0dot25_EnergySurf.png}
    \caption{Model I: the energy surface $\Sigma_{-}(h)$ for h = 0.25. }
  \end{subfigure}
  \begin{subfigure}[b]{0.42\linewidth}
    \includegraphics[width=\linewidth]{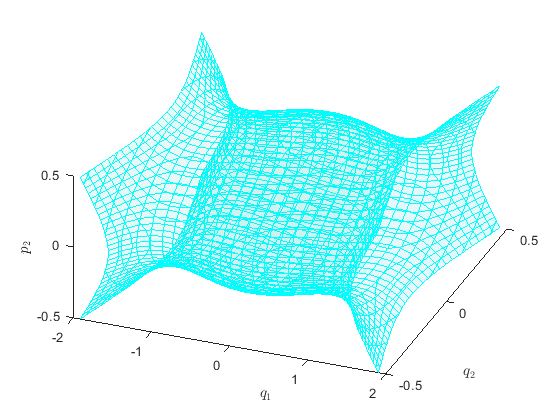}
    \caption{Model I: the energy surface $\Sigma_{+}(h)$ for h = 1. }
  \end{subfigure}
  \begin{subfigure}[b]{0.42\linewidth}
    \includegraphics[width=\linewidth]{Advanced_alpha=minus1_beta=minus1_omega=10_h=1_EnergySurf.png}
    \caption{Model I: the energy surface $\Sigma_{-}(h)$ for h = 1. }
  \end{subfigure}
  \caption{Model I: the energy surface $\Sigma_{+,-}(h)$ for h = 0, 0.25 and 1. $\alpha, \beta$, $\omega$ are chosen to be -1, -1, 10. }
  \label{fig:advancedenergy surface_alpha=-1_beta=-1}
\end{figure}

\begin{figure}[h!]
  \centering
  \begin{subfigure}[b]{0.42\linewidth}
    \includegraphics[width=\linewidth]{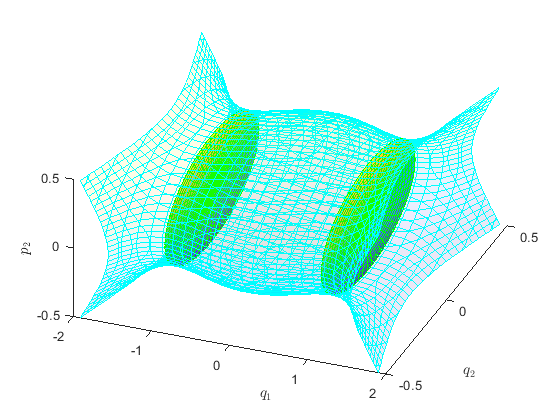}
    \caption{Model I: northern hemisphere $ B^2_{ds,f}(h)$ of DS(green) which is contained in the energy surface $\Sigma_{+}(h)$(cyan). }
  \end{subfigure}
  \begin{subfigure}[b]{0.42\linewidth}
    \includegraphics[width=\linewidth]{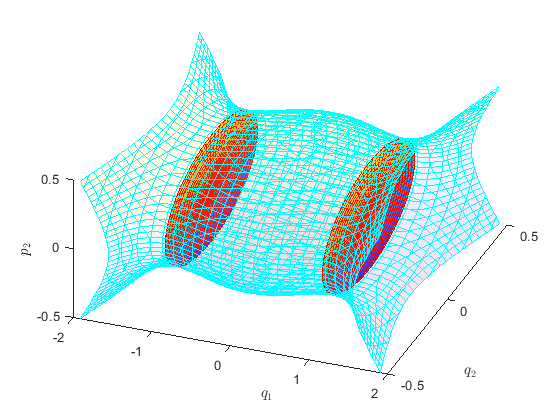}
    \caption{Model I: southern hemisphere $ B^2_{ds,b}(h)$ of DS(red) which is contained in the energy surface $\Sigma_{-}(h)$(cyan). }
  \end{subfigure}
   \begin{subfigure}[b]{0.42\linewidth}
    \includegraphics[width=\linewidth]{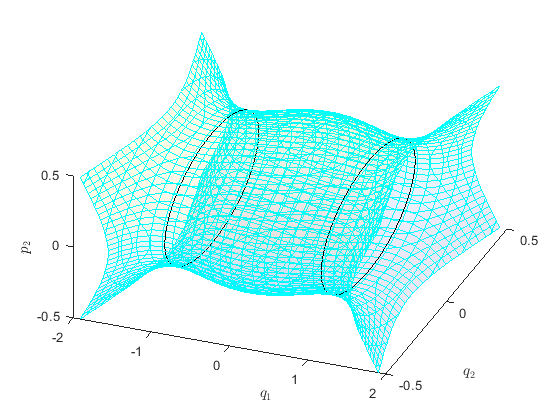}
    \caption{Model I: NHIM(black) which is contained in the energy surface $\Sigma_{+}(h)$(cyan). }
  \end{subfigure}
   \begin{subfigure}[b]{0.42\linewidth}
    \includegraphics[width=\linewidth]{Advanced_alpha=minus1_beta=minus1_omega=10_h=1_EnergySurf_NHIM.png}
    \caption{Model I: NHIM(black) which is contained in the energy surface $\Sigma_{-}(h)$(cyan). }
  \end{subfigure}
  \caption{Model I: the dividing surface(DS) and the NHIM. Here the energy surface $\Sigma_{+,-}(h)$ for h=1. $\alpha, \beta$, $\omega$ are chosen to be -1, -1, 10. }
  \label{fig:advanced DS_NHIM_alpha=-1_beta=-1}
\end{figure}

First, we present the structure of the energy surface with 3 different $h$ in Fig.~\ref{fig:advancedenergy surface_alpha=-1_beta=-1}. We present the NHIMs and the DSs in Fig.~\ref{fig:advanced DS_NHIM_alpha=-1_beta=-1}. It's interesting to see that we have 2 DSs at $ q_1 = \pm \sqrt{\alpha/\beta}$ and 2 NHIMs associated to each DS. We choose the region $q_1<-\sqrt{\alpha/\beta}$ to represent the reactants region and $q_1> \sqrt{\alpha/\beta}$ to represent the products region. The region $ -\sqrt{\alpha/\beta} <q_1 < \sqrt{\alpha/\beta}$ can be considered as an intermediate region between reactants and products. We can see from the graphs that in order for a reaction to occur, we must have trajectories crossing both DSs and this correspond to $h > -\frac{\alpha^2}{4\beta} >0$. When $ h \leq -\frac{\alpha^2}{4\beta}$, no trajectories cross the DSs. 

Note that the DSs are the intersection of the energy surface with the plane \{ $q_1 = \pm \sqrt{\alpha/\beta}$ \} in Fig.~\ref{fig:advanced DS_NHIM_alpha=-1_beta=-1}a,b and the NHIMs form the equators of the DSs in Fig.~\ref{fig:advanced DS_NHIM_alpha=-1_beta=-1}c,d. Note also that the NHIMs exist on the boundary of the energy surface.

\begin{figure}[h!]
  \centering
  \begin{subfigure}[b]{0.42\linewidth}
    \includegraphics[width=\linewidth]{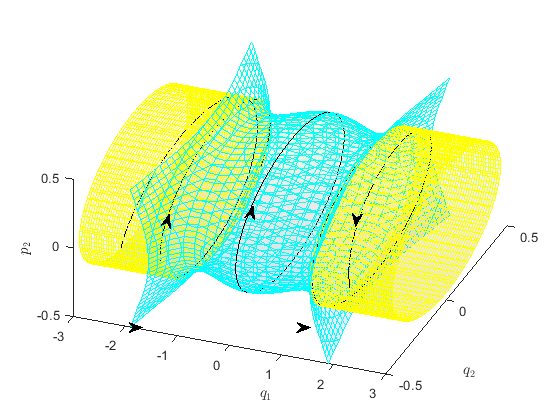}
    \caption{Model I: nonreactive trajectory(black) that is contained in invariant manifolds(yellow) $\Lambda_{I,J}(h)$ for $I= 0$, $J= 1$. Trajectory(left) is contained in the reactants component, trajectory(middle) is contained in the intermediate component and trajectory(right) is contained in the products component. Invariant manifold $\Lambda_{I,J}(h)$ are contained in $\Sigma_{+}(h)$(cyan). }
  \end{subfigure}
  \begin{subfigure}[b]{0.42\linewidth}
    \includegraphics[width=\linewidth]{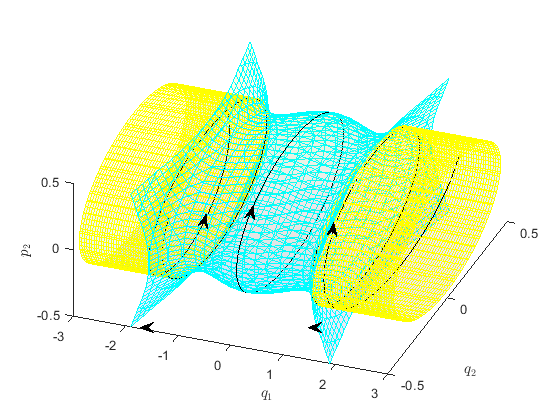}
    \caption{Model I: nonreactive trajectory(black) that is contained in a invariant manifold(yellow) $\Lambda_{I,J}(h)$ for $I= 0$, $J= 1$. Trajectory(left) is contained in the reactants component, trajectory(middle) is contained in the intermediate component and trajectory(right) is contained in the products component. Invariant manifolds $\Lambda_{I,J}(h)$ are contained in $\Sigma_{-}(h)$(cyan).}
  \end{subfigure}
  \caption{Model I: nonreactive trajectories(black). Here $I= 0$, $J= 1$. $\alpha, \beta$, $\omega$ are chosen to be -1, -1, 10. Direction of the trajectories are shown by black arrows. The arrows outside the energy surface represent an overall direction of the trajectories.}
  \label{fig: Advanced_alpha=-1_beta=-1_omega=10_I=minus0dot25_J=1dot25_LagCylin_Traj}
\end{figure}

In Fig.~\ref{fig: Advanced_alpha=-1_beta=-1_omega=10_I=minus0dot25_J=1dot25_LagCylin_Traj}, we present 3 nonreactive trajectories that are contained in $\Lambda_{I,J}(h)$ for $ I =0$. The invariant manifold consists of three separate parts and is contained in $\Sigma_{+}(h)$(cyan). The left part is in the reactants component, the middle part is in the intermediate component and the right part is in the products component. The middle part in $q_1 - p_1$ plane corresponds to the equilibrium point in the origin and therefore is a periodic orbit in $q_1 - q_2-p_2$ plane. 

\begin{figure}[h!]
\centering
\begin{subfigure}[b]{0.42\linewidth}
\includegraphics[width=\linewidth]{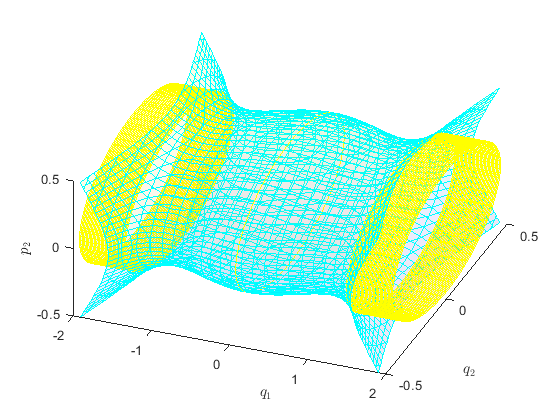}
\caption{Model I: Invariant manifolds $\Lambda_{I,J}(h)$ are contained in $\Sigma_{+}(h)$(cyan). Invariant manifold (left) is contained in the reactants component, invariant manifold(middle) is contained in the intermediate component and invariant manifold(right) is contained in the products component.}
\end{subfigure}
\begin{subfigure}[b]{0.42\linewidth}
\includegraphics[width=\linewidth]{Advanced_alpha=minus1_beta=minus1_omega=10_I=0_J=1_LagCylin.png}
\caption{Model I: Invariant manifolds $\Lambda_{I,J}(h)$ are contained in $\Sigma_{-}(h)$(cyan). Invariant manifold (left) is contained in the reactants component, invariant manifold(middle) is contained in the intermediate component and invariant manifold(right) is contained in the products component.}
\end{subfigure}
\caption{Model I: Invariant manifolds(yellow). Here $I= 0$, $J= 1$. $\alpha, \beta$, $\omega$ are chosen to be -1, -1, 10.}
\label{fig: Advanced_alpha=-1_beta=-1_omega=10_I=minus0dot25_J=1dot25_LagCylin}
\end{figure}

For clarity, we present the invariant manifolds separately in Fig.~\ref{fig: Advanced_alpha=-1_beta=-1_omega=10_I=minus0dot25_J=1dot25_LagCylin}. 

In Fig.~\ref{fig: Advanced_alpha=-1_beta=-1_omega=10_I=minus0dot25_J=1dot25_LagCylin_Traj}a, left, we present a trajectory which starts from the reactants component, tries to approach the DS($q_1 = -\sqrt{\alpha/\beta}$), it reaches the boundary of the energy surface$\Sigma_{+}(h)$(cyan), where $ p_1 =0$ before the DS and jumps to $\Sigma_{-}(h)$(cyan) in Fig.~\ref{fig: Advanced_alpha=-1_beta=-1_omega=10_I=minus0dot25_J=1dot25_LagCylin_Traj}b, left. It then returns back to the reactants component with $p_1 < 0$. 

In Fig.~\ref{fig: Advanced_alpha=-1_beta=-1_omega=10_I=minus0dot25_J=1dot25_LagCylin_Traj}a,b, middle, we present a trajectory which starts on the periodic orbit, stays on the periodic orbit for all time. Note that the invariant manifold in this case exists on the boundary of the energy surface and with $p_1 = 0$ for all time. 

In Fig.~\ref{fig: Advanced_alpha=-1_beta=-1_omega=10_I=minus0dot25_J=1dot25_LagCylin_Traj}b, right, we present a trajectory which starts from the products component, tries to approach the DS($q_1 = \sqrt{\alpha/\beta}$). It reaches the boundary of the energy surface$\Sigma_{-}(h)$(cyan), where $ p_1 =0$ before the DS and jumps to $\Sigma_{+}(h)$(cyan) in Fig.~\ref{fig: Advanced_alpha=-1_beta=-1_omega=10_I=minus0dot25_J=1dot25_LagCylin_Traj}a, right. It then returns back to the products component with $p_1 > 0$

\begin{figure}[h!]
  \centering
  \begin{subfigure}[b]{0.42\linewidth}
    \includegraphics[width=\linewidth]{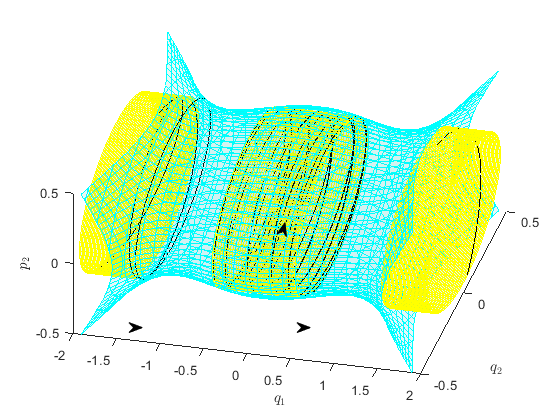}
    \caption{Model I: nonreactive trajectory(black) that is contained in invariant manifolds(yellow) $\Lambda_{I,J}(h)$ for $I= 0.1$, $J= 0.9$. Trajectory(left) is contained in the reactants component, trajectory(middle) is contained in the intermediate component and trajectory(right) is contained in the products component. Invariant manifolds $\Lambda_{I,J}(h)$ are contained in $\Sigma_{+}(h)$(cyan). The arrows outside the Energy surface represent the trajectories oscillate from left to right. }
  \end{subfigure}
  \begin{subfigure}[b]{0.42\linewidth}
    \includegraphics[width=\linewidth]{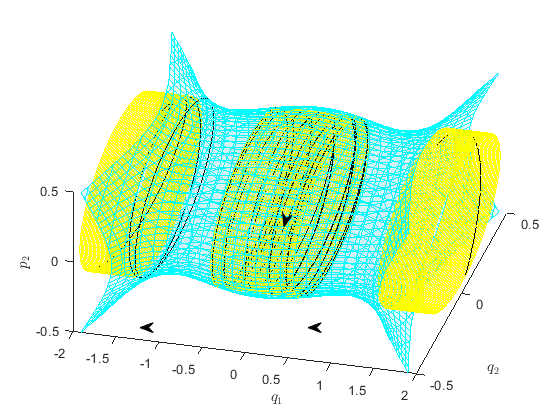}
    \caption{Model I: nonreactive trajectory(black) that is contained in invariant manifolds(yellow) $\Lambda_{I,J}(h)$ for $I= 0.1$, $J= 0.9$. Trajectory(left) is contained in the reactants component, trajectory(middle) is contained in the intermediate component and trajectory(right) is contained in the products component. Invariant manifolds $\Lambda_{I,J}(h)$ are contained in $\Sigma_{-}(h)$(cyan). The arrows outside the Energy surface represent the trajectories go from right to left. }
  \end{subfigure}
  \caption{Model I: nonreactive trajectories(black). Here $I= 0.1$, $J= 0.9$. $\alpha, \beta$, $\omega$ are chosen to be -1, -1, 10. Direction of the trajectories are shown by black arrows. The arrows outside the energy surface represent an overall direction of the trajectories.}
  \label{fig: Advanced_alpha=-1_beta=-1_omega=10_I=0dot1_J=0dot9_LagCylin_Traj_forward}
\end{figure}

In Fig.~\ref{fig: Advanced_alpha=-1_beta=-1_omega=10_I=0dot1_J=0dot9_LagCylin_Traj_forward}a, we present 3 nonreactive trajectories that are contained in $\Lambda_{I,J}(h)$ for $ 0 < I < \frac{\alpha^2}{4\beta} $. The invariant manifold consists of 3 separate parts and is contained in $\Sigma_{+}(h)$(cyan). The left part is in the reactants component, the middle part is in the intermediate component and the right part is in the products component. 

In Fig.~\ref{fig: Advanced_alpha=-1_beta=-1_omega=10_I=0dot1_J=0dot9_LagCylin_Traj_forward}b, we present 3 nonreactive trajectories that are contained in $\Lambda_{I,J}(h)$ for $ 0 < I < \frac{\alpha^2}{4\beta} $. The invariant manifold consists of 3 separate parts and is contained in $\Sigma_{-}(h)$(cyan). The left part is in the reactants component, the middle part is in the intermediate component and the right part is in the products component. 

The direction of the trajectories are determined in the following: 

Suppose that a  trajectory starts on the invariant manifold in the reactants component(left part of the invariant manifold in Fig.~\ref{fig: Advanced_alpha=-1_beta=-1_omega=10_I=0dot1_J=0dot9_LagCylin_Traj_forward}a, it goes to the right until it reaches the boundary of the energy surface $\Sigma_{+}(h)$, where $p_1=0$ and it jumps to $\Sigma_{-}(h)$(left part in Fig.~\ref{fig: Advanced_alpha=-1_beta=-1_omega=10_I=0dot1_J=0dot9_LagCylin_Traj_forward}b. It then returns to the reactants region and no reaction occurs.
Similarly, if the trajectory starts on the invariant manifold in the products component(right part of the invariant manifold in Fig.~\ref{fig: Advanced_alpha=-1_beta=-1_omega=10_I=0dot1_J=0dot9_LagCylin_Traj_forward}b, it goes to the left until it reaches the boundary of the energy surface $\Sigma_{-}(h)$, where $p_1=0$ and it jumps to $\Sigma_{+}(h)$(right part in Fig.~\ref{fig: Advanced_alpha=-1_beta=-1_omega=10_I=0dot1_J=0dot9_LagCylin_Traj_forward}a. It then returns to the reactants region and no reaction occurs. 

In Fig.~\ref{fig: Advanced_alpha=-1_beta=-1_omega=10_I=0dot1_J=0dot9_LagCylin_Traj_forward}a, middle we present a trajectory which starts on the invariant manifold in the intermediate component, it oscillates to the right until it reaches the boundary of the energy surface $\Sigma_{+}(h)$(right end of the invariant manifold in the middle), where $p_1=0$ and it jumps to $\Sigma_{-}(h)$(right end of the invariant manifold in the middle in Fig.~\ref{fig: Advanced_alpha=-1_beta=-1_omega=10_I=0dot1_J=0dot9_LagCylin_Traj_forward}b. It then oscillates to the left until it reaches the boundary of the energy surface $\Sigma_{-}(h)$(left end of the invariant manifold in the middle), where $p_1=0$ and it jumps back to $\Sigma_{+}(h)$(left end of the invariant manifold in the middle in Fig.~\ref{fig: Advanced_alpha=-1_beta=-1_omega=10_I=0dot1_J=0dot9_LagCylin_Traj_forward}a. It then oscillates to the right and we have the same motion again. No reaction occurs in this case as the trajectory oscillates back before it reaches the NHIM. Any trajectories start in the reactants/products component stay in the reactants/products component for all time. Any trajectories start in the intermediate component oscillate in the intermediate component for all time. The overall directions of the trajectories are shown by the arrows outside the energy surface.

\begin{figure}[h!]
  \centering
  \begin{subfigure}[b]{0.42\linewidth}
    \includegraphics[width=\linewidth]{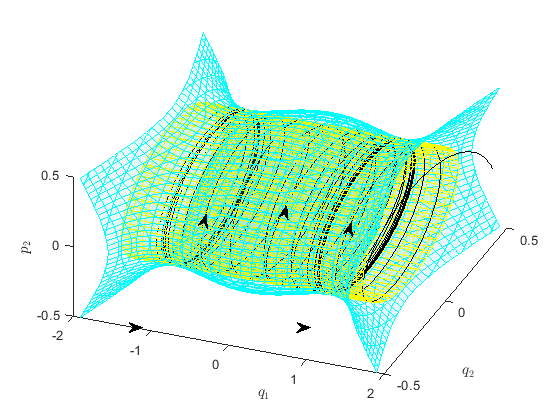}
    \caption{Model I: forward branches of the NHIMs(yellow) are contained in $\Sigma_{+}(h)$(cyan). }
  \end{subfigure}
  \begin{subfigure}[b]{0.42\linewidth}
    \includegraphics[width=\linewidth]{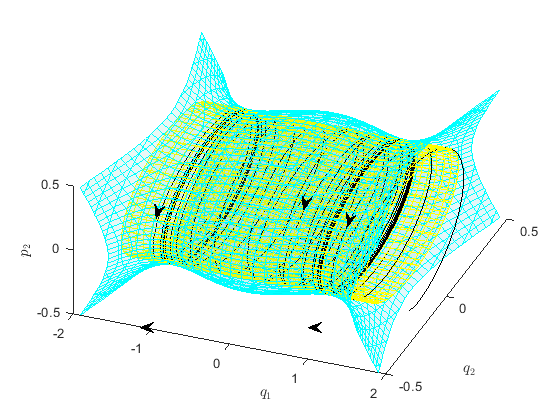}
    \caption{Model I: backward branches of the NHIMs(yellow) that are contained in $\Sigma_{-}(h)$(cyan). }
  \end{subfigure}
  \caption{Model I: stable and unstable manifolds of the NHIM. Here $I = 0.25$, $J= 0.75$. $\alpha, \beta$, $\omega$ are chosen to be -1, -1, 10. Direction of the trajectories are shown by black arrows. The arrows outside the energy surface represent an overall direction of the trajectories.}
  \label{Advanced_alpha=-1_beta=-1_omega=10_I=0.25_J=0.75_LagCylin_Traj_Sta_Unsta}
\end{figure}

We present the stable and unstable manifolds of the NHIMs in Fig.~\ref{Advanced_alpha=-1_beta=-1_omega=10_I=0.25_J=0.75_LagCylin_Traj_Sta_Unsta}. \\
In Fig.~\ref{Advanced_alpha=-1_beta=-1_omega=10_I=0.25_J=0.75_LagCylin_Traj_Sta_Unsta}a left, the trajectory which starts on the $W^{s}_{b}(-\sqrt{\alpha/\beta},0)$(reactants region) approaches the NHIM
\begin{align}
     \mathbb{S}^{1}_{\text{NHIM}}(h) = \Sigma(h) \cap \{ q_1 = - \sqrt{\alpha/\beta}, p_1 = 0 \} = \{ (q_1, p_1, q_2, p_2 ) \in \mathbb{R}^4 : \omega(p_2^2 + q_2^2) = 2h +\frac{\alpha^2}{2\beta} \} ,\\ q_1 = - \sqrt{\alpha/\beta}, p_1 = 0.
\end{align}
as $ t \rightarrow \infty$. In Fig.~\ref{Advanced_alpha=-1_beta=-1_omega=10_I=0.25_J=0.75_LagCylin_Traj_Sta_Unsta}b, left The trajectory which starts on the $W^{u}_{b}(-\sqrt{\alpha/\beta},0)$(products region) approaches the NHIM
\begin{align}
     \mathbb{S}^{1}_{\text{NHIM}}(h) = \Sigma(h) \cap \{ q_1 = - \sqrt{\alpha/\beta}, p_1 = 0 \} = \{ (q_1, p_1, q_2, p_2 ) \in \mathbb{R}^4 : \omega(p_2^2 + q_2^2) = 2h +\frac{\alpha^2}{2\beta} \} ,\\ q_1 = - \sqrt{\alpha/\beta}, p_1 = 0.
\end{align}
as $ t \rightarrow -\infty$. \\
In Fig.~\ref{Advanced_alpha=-1_beta=-1_omega=10_I=0.25_J=0.75_LagCylin_Traj_Sta_Unsta}b, right, the trajectory which starts on the $W^{s}_{b}(\sqrt{\alpha/\beta},0)$(productss region) approaches the NHIM
\begin{align}
     \mathbb{S}^{1}_{\text{NHIM}}(h) = \Sigma(h) \cap \{ q_1 = + \sqrt{\alpha/\beta}, p_1 = 0 \} = \{ (q_1, p_1, q_2, p_2 ) \in \mathbb{R}^4 : \omega(p_2^2 + q_2^2) = 2h +\frac{\alpha^2}{2\beta} \} ,\\ q_1 = + \sqrt{\alpha/\beta}, p_1 = 0.
\end{align}
as $ t \rightarrow \infty$. In Fig.~\ref{Advanced_alpha=-1_beta=-1_omega=10_I=0.25_J=0.75_LagCylin_Traj_Sta_Unsta}a, right, the trajectory which starts on the $W^{u}_{b}(\sqrt{\alpha/\beta},0)$(products region) approaches the NHIM
\begin{align}
     \mathbb{S}^{1}_{\text{NHIM}}(h) = \Sigma(h) \cap \{ q_1 = + \sqrt{\alpha/\beta}, p_1 = 0 \} = \{ (q_1, p_1, q_2, p_2 ) \in \mathbb{R}^4 : \omega(p_2^2 + q_2^2) = 2h +\frac{\alpha^2}{2\beta} \} ,\\ q_1 = + \sqrt{\alpha/\beta}, p_1 = 0.
\end{align}
as $ t \rightarrow -\infty$. \\
In Fig.~\ref{Advanced_alpha=-1_beta=-1_omega=10_I=0.25_J=0.75_LagCylin_Traj_Sta_Unsta}a, middle , trajectory approaches the NHIM
\begin{align}
     \mathbb{S}^{1}_{\text{NHIM}}(h) = \Sigma(h) \cap \{ q_1 = - \sqrt{\alpha/\beta}, p_1 = 0 \} = \{ (q_1, p_1, q_2, p_2 ) \in \mathbb{R}^4 : \omega(p_2^2 + q_2^2) = 2h +\frac{\alpha^2}{2\beta} \} ,\\ q_1 = - \sqrt{\alpha/\beta}, p_1 = 0.
\end{align}
as $ t \rightarrow -\infty$ and approaches the 
NHIM
\begin{align}
     \mathbb{S}^{1}_{\text{NHIM}}(h) = \Sigma(h) \cap \{ q_1 = + \sqrt{\alpha/\beta}, p_1 = 0 \} = \{ (q_1, p_1, q_2, p_2 ) \in \mathbb{R}^4 : \omega(p_2^2 + q_2^2) = 2h +\frac{\alpha^2}{2\beta} \} ,\\ q_1 = + \sqrt{\alpha/\beta}, p_1 = 0.
\end{align}
as $ t \rightarrow \infty$. Note that the trajectory stay in the intermediate region for all time.\\
In Fig.~\ref{Advanced_alpha=-1_beta=-1_omega=10_I=0.25_J=0.75_LagCylin_Traj_Sta_Unsta}b, middle , trajectory approaches the NHIM
\begin{align}
     \mathbb{S}^{1}_{\text{NHIM}}(h) = \Sigma(h) \cap \{ q_1 = - \sqrt{\alpha/\beta}, p_1 = 0 \} = \{ (q_1, p_1, q_2, p_2 ) \in \mathbb{R}^4 : \omega(p_2^2 + q_2^2) = 2h +\frac{\alpha^2}{2\beta} \} ,\\ q_1 = - \sqrt{\alpha/\beta}, p_1 = 0.
\end{align}
as $ t \rightarrow \infty$ and approaches the 
NHIM 
\begin{align}
     \mathbb{S}^{1}_{\text{NHIM}}(h) = \Sigma(h) \cap \{ q_1 = + \sqrt{\alpha/\beta}, p_1 = 0 \} = \{ (q_1, p_1, q_2, p_2 ) \in \mathbb{R}^4 : \omega(p_2^2 + q_2^2) = 2h +\frac{\alpha^2}{2\beta} \} ,\\ q_1 = + \sqrt{\alpha/\beta}, p_1 = 0.
\end{align}
as $ t \rightarrow -\infty$. Note that the trajectories stay in the intermediate region for all time.

\begin{figure}[h!]
  \centering
  \begin{subfigure}[b]{0.42\linewidth}
    \includegraphics[width=\linewidth]{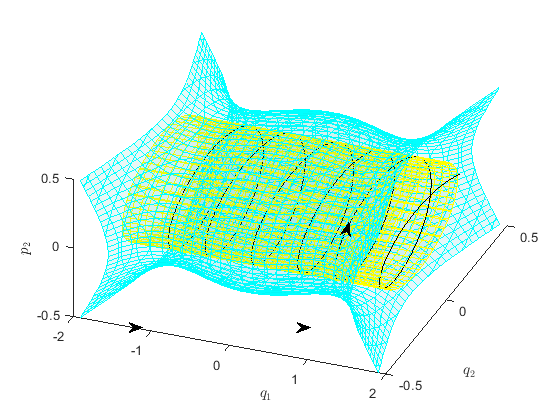}
    \caption{Model I: forward reactive trajectory(black) that is contained in a invariant manifold(yellow) $\Lambda_{I,J}(h)$ for $I= 0.5$, $J= 0.5$. Invariant manifold $\Lambda_{I,J}(h)$ is contained in $\Sigma_{+}(h)$(cyan). }
  \end{subfigure}
  \begin{subfigure}[b]{0.42\linewidth}
    \includegraphics[width=\linewidth]{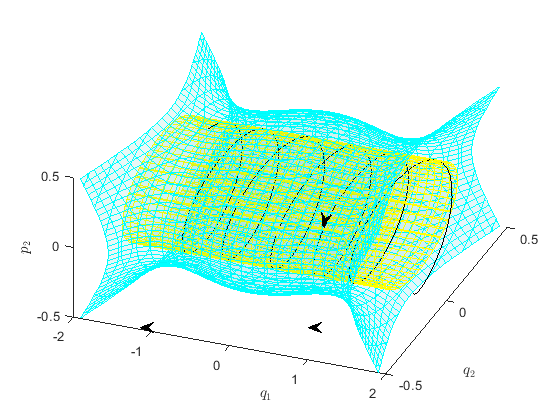}
    \caption{Model I: Backward reactive trajectory(black) that is contained in a invariant manifold(yellow) $\Lambda_{I,J}(h)$ for $I= 0.5$, $J= 0.5$. Invariant manifold $\Lambda_{I,J}(h)$ is contained in $\Sigma_{-}(h)$(cyan). }
  \end{subfigure}
  \caption{Model I: reactive trajectories contained in the invariant manifolds. Here $I = 0.5$, $J= 0.5$. $\alpha, \beta$, $\omega$ are chosen to be -1, -1, 10. Direction of the trajectories are shown by black arrows.}
  \label{Advanced_alpha=-1_beta=-1_omega=10_I=0dot5_J=0dot5_LagCylin_Traj}
\end{figure}

In Fig.~\ref{Advanced_alpha=-1_beta=-1_omega=10_I=0dot5_J=0dot5_LagCylin_Traj} we present a forward in Fig.~\ref{Advanced_alpha=-1_beta=-1_omega=10_I=0dot5_J=0dot5_LagCylin_Traj}a and a backward in Fig.~\ref{Advanced_alpha=-1_beta=-1_omega=10_I=0dot5_J=0dot5_LagCylin_Traj}b reactive trajectory that are contained in invariant manifolds. In Fig.~\ref{Advanced_alpha=-1_beta=-1_omega=10_I=0dot5_J=0dot5_LagCylin_Traj}a, the trajectory comes from reactants region($q_1 < - \sqrt{\alpha/\beta}$), goes through the DS($q_1= - \sqrt{\alpha/\beta}$) and then enters intermediate region($- \sqrt{\alpha/\beta}<q_1 <\sqrt{\alpha/\beta}$). It then goes through the other DS($ q_1 = \sqrt{\alpha/\beta}$), enters the products region($ q_1 > \sqrt{\alpha/\beta}$). It's therefore characterised as a forward reactive trajectory. 

In Fig.~\ref{Advanced_alpha=-1_beta=-1_omega=10_I=0dot5_J=0dot5_LagCylin_Traj}b, the trajectory comes from products region($q_1 >  \sqrt{\alpha/\beta}$), goes through the DS($q_1=  \sqrt{\alpha/\beta}$) and then enters intermediate region($- \sqrt{\alpha/\beta}<q_1 <\sqrt{\alpha/\beta}$). It then goes through the other DS($ q_1 = -\sqrt{\alpha/\beta}$), enters the reactants region($ q_1 < -\sqrt{\alpha/\beta}$). It's therefore characterised as a backward reactive trajectory. \\
In both cases the trajectory enters the DS twice and it therefore reacts twice.

\subsection{Numerical methods for computing unstable periodic orbits\label{subsect:numeric_methods}}

In this section, we discuss turning point and tuning point based on configuration difference methods for computing the NHIM. As our Hamiltonian system is integrable and there is an analytical form for the NHIM, we can always compare the result obtained using different methods with the analytical solution to see whether the two results agreed with each other or not. Or we can calculate the Hausdorff distance~\cite{Hausdorff1914} between the UPOs obtained using a numerical method and analytical solution for the uncoupled system which is a more rigorous way of checking the convergence.

\subsubsection{Turning point method\label{subsubsect:turning_pt}}

Note first that we can rewrite the Hamiltonian in \eqref{defn:hamiltonian} as kinetic energy $T(p_1,p_2)$ plus potential energy $V(q_1,q_2)$ and the $q_1-q_2$ plane is the usual configuration plane.
\begin{align} H = T(p_1,p_2)+V(q_1,q_2) = \frac{ 1}{2} p_1^2 + \frac{\omega}{2}p_2^2 - \frac{\alpha  }{2}q_1^2+ \frac{\beta }{4}q_1^4 + \frac{\omega}{2} q_2 ^2
\end{align} 

Suppose that we have 2 trajectories start on the same side of the equipotential contour $V(q_1,q_2) = e$ where $V(q_1,q_2)$ is the potential energy function and $e$ is the total energy of the system. Note that this means that our kinetic energy at the initial position is zero since $H=T(p_1,p_2)+V(q_1,q_2) $ which implies that $T+e=e$. Let us denote the starting position of the two trajectories by $(q_{1,1}^0,q_{2,1}^0, p_{1,1}^0, p_{2,1}^0)$, $ (q_{1,2}^0,q_{2,2}^0, p_{1,2}^0, p_{2,2}^0)$. If the two starting positions are close to each other, the two trajectories follow a closely path until they reach their first turning points, denote by $(q_{1,1}^{\text{tp}1},q_{2,1}^{\text{tp}1}, p_{1,1}^{\text{tp}1}, p_{2,1}^{\text{tp}1})$, $ (q_{1,2}^{\text{tp}1},q_{2,2}^{\text{tp}1}, p_{1,2}^{\text{tp}1}, p_{2,2}^{\text{tp}1})$. When we talk about turning points, we refer to the points where either $p_1$ or $p_2$ coordinate of the trajectory becomes zero and require the same momentum to be zero for all turning points. Note that the first turning point is on the different side of the equipotential contour. After the trajectories reach their first turning points, the momentum used to define the turning points changes its sign. The trajectories bounce back from the different side of the equipotential contour $V(x,y) = e$ and move back to the initial side of the equipotential contour until they reach their second turning points $(q_{1,1}^{\text{tp}2},q_{2,1}^{\text{tp}2}, p_{1,1}^{\text{tp}2}, p_{2,1}^{\text{tp}2})$, $ (q_{1,2}^{\text{tp}2},q_{2,2}^{\text{tp}2}, p_{1,2}^{\text{tp}2}, p_{2,2}^{\text{tp}2})$ and the motion continues. Note that if a trajectory starts on the NHIM and move to the other side of the equipotential contour, both values of $p_1$ and $p_2$ become $0$ at their turning points, i.e. the turning points for $p_1$ is the same as the turning points for $p_2$. 
\begin{align*}
    (q_{1}^{\text{tp}1},q_{2}^{\text{tp}1}, p_{1}^{\text{tp}1}, p_{2}^{\text{tp}1}) _=(q_{1}^{\text{tp}1},q_{2}^{\text{tp}1}, 0,0) \\
    (q_{1}^{\text{tp}2},q_{2}^{\text{tp}2}, p_{1}^{\text{tp}2}, p_{2}^{\text{tp}2}) &=(q_{1}^{\text{tp}2},q_{2}^{\text{tp}2}, 0,0) \\
    &...
\end{align*}
The trajectories are obtained by integrating the Hamilton's equations in (\ref{newhamq_1}),(\ref{newhamp_1}),(\ref{newhamq_2}),(\ref{newhamp_2}) with respect to the initial conditions on the equipotential line where $p_{1,1}^0= p_{2,1}^0=p_{1,2}^0= p_{2,2}^0=0$. When we use computers to perform numerical integration, time is discretised and we can record the time when the trajectory is at the turning point using event integration or an approximate time by checking the signs of the momentum. Without loss of generality, we assume the discretised time interval for integration is $[t_0, t_1, ..., t_n]$. For example, if a trajectory has a positive value of $p_1$ at time $t_i, i \in \{0, ..., n\} $ but a negative value of $p_1$ at time $t_{i+1}$, we know the trajectory has reached its turning point during the time interval $[t_i,t_{i+1}]$. The time $t_i$ is an approximate time for turning and the position of the trajectory at time $t_i$ is an approximate position of the turning point.

\begin{figure}[!ht]
	\centering
	\includegraphics[width=0.49\textwidth]{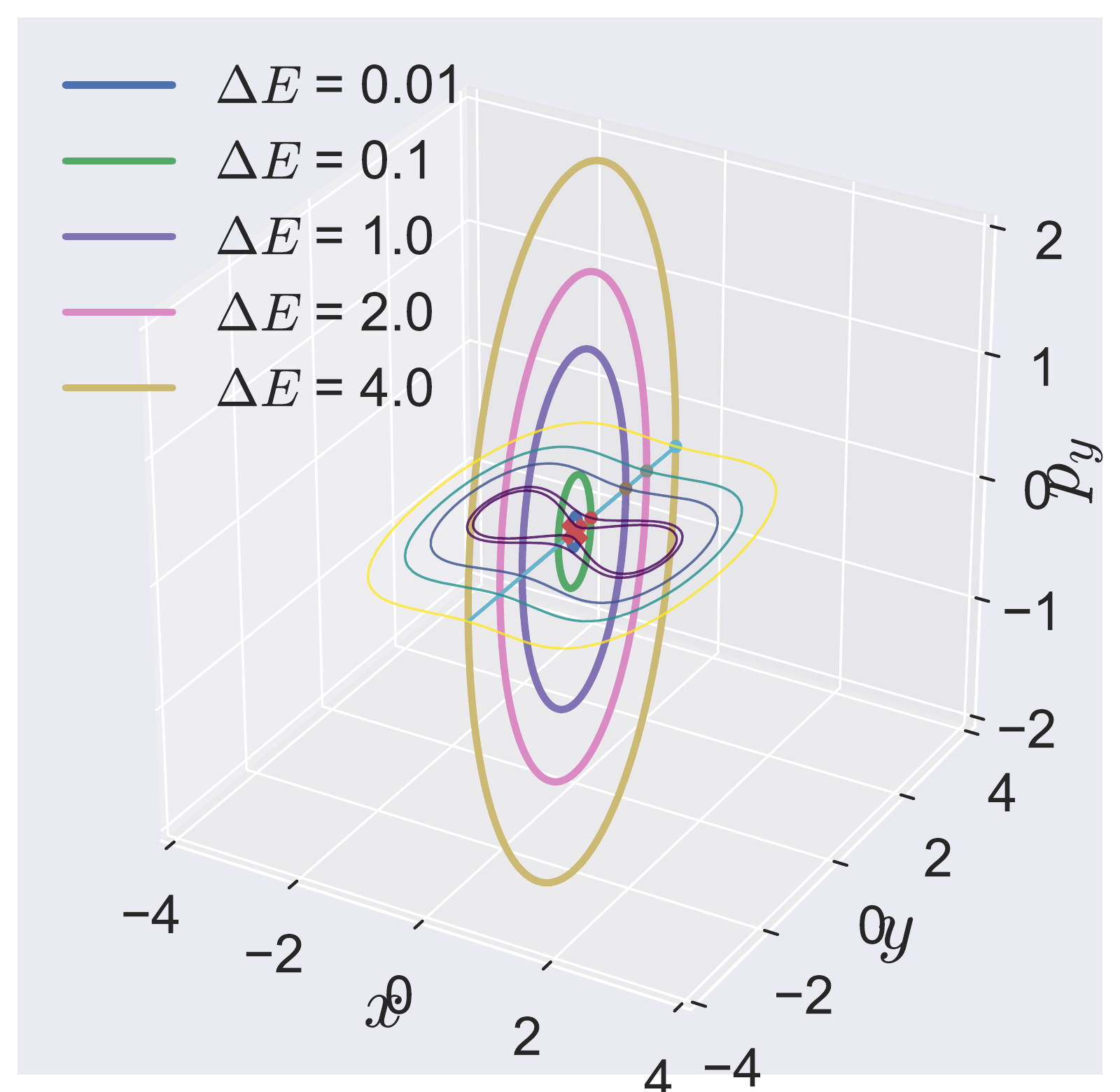}
	\caption{\textbf{Family of NHIMs computed using turning point method with equipotential lines in the $x-y$ plane.} Parameters are $\alpha=1,\beta=1,\omega=1$.}\label{fig: UPO_uncoupled_tp}
\end{figure}

\begin{figure}[!ht]
	\centering
	\includegraphics[width=0.49\textwidth]{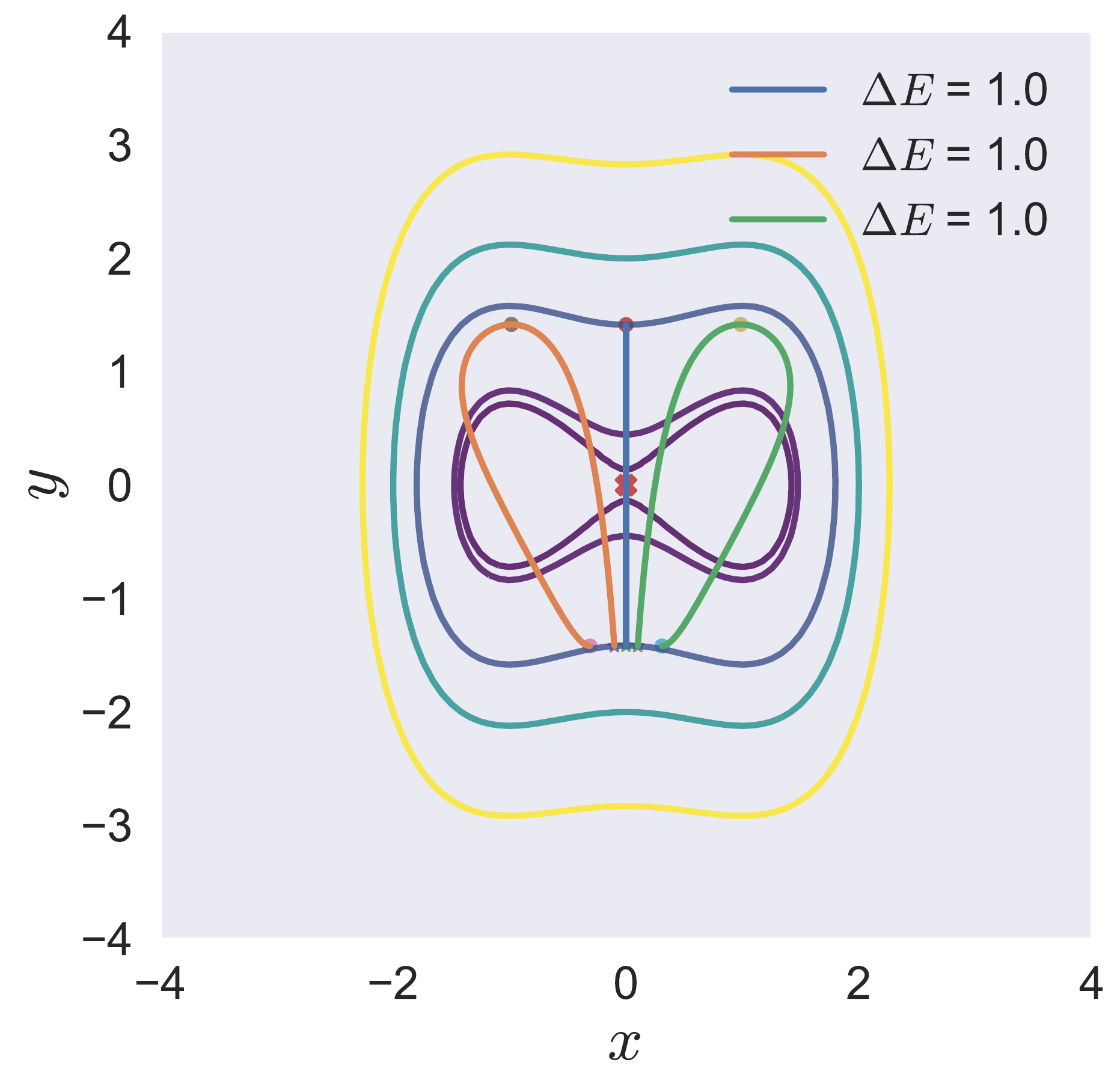}
	\caption{\textbf{Illustration of the turning point method.} NHIM is shown in blue and two representative trajectories that turn in opposite directions are shown in orange and green. The starting points are marked as $\star$ and the turning points are marked as $\bullet$. Parameters are $\alpha=1,\beta=1,\omega=1$.}\label{fig: UPO_illustration}
\end{figure}

In principle, we can check the signs of $p_{1,1}^{\text{tp}1} \cdot p_{1,2}^{\text{tp}1}$ and $p_{2,1}^{\text{tp}1} \cdot p_{2,2}^{\text{tp}1}$. If the signs are the same and positive, both trajectories turn in the same direction. If the signs are different and one of the signs is negative, both trajectories turn in opposite directions. If the trajectories turn in opposite directions, the NHIM must exist between the two trajectories. We can shorten the distance between the initial conditions $(q_{1,1}^{0},q_{2,1}^{0})$ and $(q_{1,2}^{0},q_{2,2}^{0})$ and use the principle again. However, this condition is sensitive to the position of the turning point we use, therefore let's define the following 
\begin{align*}
    p_{1,i}^{ \perp} &= |(p_{1,i}(t_{ i-2}),p_{1,i}(t_{ i-2}))| p_{1,i} (t_{ i-1}) - [(p_{1,i} (t_{ i-1}),p_{2,i} (t_{ i-1})) \cdot (p_{1,i} (t_{ i-2}),p_{2,i} (t_{ i-2})) ]p_{1,i} (t_{i-2}) \\
    p_{2,i}^{ \perp} &= |(p_{1,i}(t_{ i-2}),p_{1,i}(t_{ i-2}))| p_{2,i} (t_{ i-1}) - [(p_{1,i} (t_{ i-1}),p_{2,i} (t_{ i-1})) \cdot (p_{1,i} (t_{ i-2}),p_{2,i} (t_{ i-2})) ]p_{2,i} (t_{i-2}) , i=1,2\\
\end{align*}
where $p_{1,i}(t),p_{2,i}(t)$ are momenta at time $t$, trajectories reach their turning points at time $t_i$ or during the time interval $[t_i,t_{t+1}]$. We use the following dot product condition to check whether the trajectories turn in the opposite directions or not.
\begin{equation*}
    (p_{1,1}^{ \perp},p_{2,1}^{ \perp}) \cdot (p_{1,2}^{ \perp},p_{2,2}^{ \perp})
\end{equation*}
In order to use this formula, we don't need to know the exact time for turning, we just need to know the position of the trajectories at one and two time steps before turning. A family of the UPOs obtained using turning point method is shown in Fig. \ref{fig: UPO_uncoupled_tp} and an illustration of position of the turning point is shown in Fig.~\ref{fig: UPO_illustration}.
\subsubsection{Turning point based on configuration difference method}
\begin{figure}[!ht]
	\centering
	\includegraphics[width=0.49\textwidth]{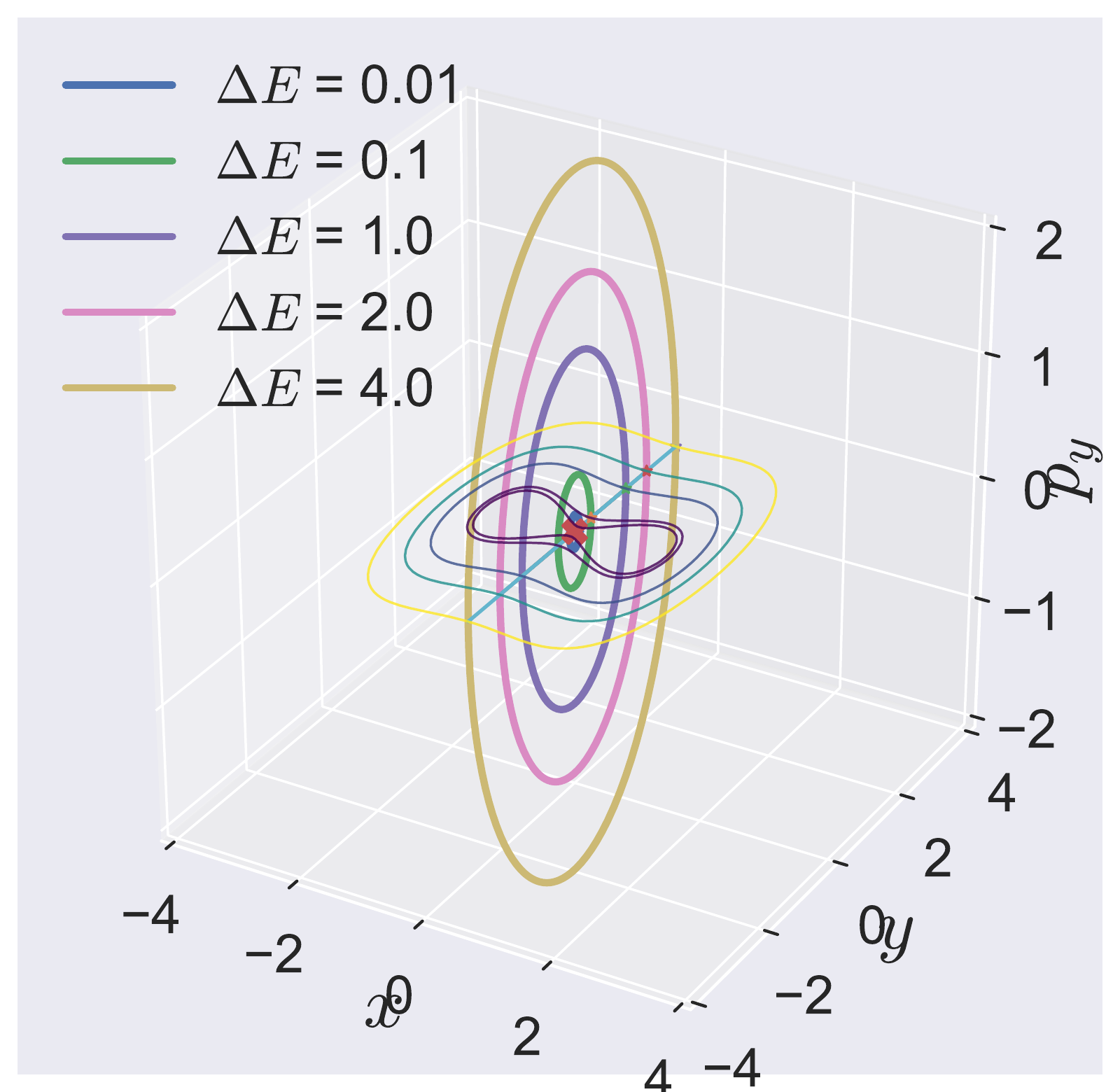}
	\caption{\textbf{Family of NHIMs computed using tpcd method with equipotential lines in the $x-y$ plane.} Parameters are $\alpha=1,\beta=1,\omega=1$. }\label{fig: UPO_uncoupled_tpcd}
\end{figure}

Base on the turning point method discussed in the previous subsection, we now discuss the turning point based on configuration difference method. The original turning point method uses a dot product condition to detect whether the trajectories turn in opposite directions and it leads to stalling of the method beyond a certain tolerance. The idea of this method is to use a different condition rather than the dot product condition to check whether the trajectories are turning in opposite directions. As described in the previous subsection, we are able to find the position of the turning point (or an approximate position of the turning point). We can then take the difference of configuration space variables $q_1$, $q_2$ between the first turning point and the starting point.
\begin{equation*}
    q_{1,\text{diff}1} = q_1^{0}-q_1^{\text{tp}1}, q_{2,\text{diff}2}  = q_2^{0}-q_2^{\text{tp}1}
\end{equation*}
We can also define the configuration space variables between any turning points and the starting point. 
\begin{equation*}
    q_{1,\text{diff}i} = q_1^{0}-q_1^{\text{tp}i}, q_{2,\text{diff}i}  = q_2^{0}-q_2^{\text{tp}i}
\end{equation*}
If we use $p_1$ to define our turning points, we want to check the configuration differences $q_{1,\text{diff}1}$ of the two trajectories. if the signs of two configuration differences are both positive, the two trajectories turn in the same direction. If one of the signs is positive and the other one is negative, the two trajectories turn in the opposite directions. The NHIM must exist between them and we can shorten the interval between the two initial conditions and apply the principle again. The idea of the method comes from the fact that if the NHIM exists between two trajectories, the trajectory on the left of the NHIM will turn to the left, the trajectory on the right of the NHIM will turn to the right. Clearly the same idea can be applied to other systems where the position of the NHIM are different. In practise, we often use $q_{1,\text{diff}2}$ and $q_{2,\text{diff}2}$ since the second turning point and the starting point are on the same side of the equipotential contour. A illustration of the UPOs obtained using turning point based on configuration difference method is shown in Fig. \ref{fig: UPO_uncoupled_tpcd}.

\section{Conclusion and Outlook\label{sect:conclusions}}

In this article, we discussed the methods for realizing the phase space structures associated with an index-1 saddle in a two degrees-of-freedom (DOF) Hamiltonian system. This Hamiltonian system can be viewed as an extension of the quadratic normal form Hamiltonian with the additional quartic term $\beta q_1^4/4$ term in our potential energy function. This quartic Hamiltonian where the two coordinates are uncoupled, and thus lets us obtain the phase space structures analytically for a nonlinear system. We showed that the changes in the sign of the parameters of the potential energy leads to Hamiltonian pitchfork bifurcation which generates two bottlenecks with their corresponding unstable periodic orbit, DS, and invariant manifolds. We also discussed the turning point and turning point based on configuration difference methods for computing the unstable periodic orbit associated with an index-1 saddle. We have shown that the UPOs obtained using both the numerical methods agreed with the analytical solutions and thus show their usefulness for general two DOF systems. This example system and the coupled system have been used in the open-source package, UPOsHam~\cite{Lyu2020}.

The future work will discuss the bifurcation of the DS in the coupled system with a quadratic coupling term in the potential energy. The numerical methods discussed here will be used in calculating the unstable periodic orbit and sampling the phase space dividing surface since the coupling makes the analytical expression infeasible.

\bibliography{references.bib}  

\begin{thebibliography}{10}

\bibitem{agaoglou_chemical_2019}
M.~Agaoglou, B.~Aguilar-Sanjuan, V.~J. Garc{\'i}a-Garrido,
  R.~Garc{\'i}a-Meseguer, F.~Gonz{\'a}lez-Montoya, M.~Katsanikas, V.~Kraj{\v
  n}{\'a}k, S.~Naik, and S.~Wiggins.
\newblock {\em {Chemical Reactions: A Journey into Phase Space}}.
\newblock Zenodo v0.1.0, Dec 2019.
\newblock \url{https://www.chemicalreactions.io} (EPSRC Grant Number:
  EP/P021123/1).

\bibitem{Dellnitz2005}
M.~Dellnitz, O.~Junge, M.~W. Lo, J.~E. Marsden, K.~Padberg, R.~Preis, S.~D.
  Ross, and B.~Thiere.
\newblock Transport of mars-crossing asteroids from the quasi-hilda region.
\newblock {\em Phys. Rev. Lett.}, 94:231102, Jun 2005.

\bibitem{Ezra2009}
G.~S. Ezra, H.~Waalkens, and S.~Wiggins.
\newblock Microcanonical rates, gap times, and phase space dividing surfaces.
\newblock {\em The Journal of Chemical Physics}, 130(16):164118, 2009.

\bibitem{farantos_pomult_1998}
S.~C. Farantos.
\newblock Pomult: A program for computing periodic orbits in hamiltonian
  systems based on multiple shooting algorithms.
\newblock {\em Computer physics communications}, 108(2-3):240--258, 1998.

\bibitem{Hausdorff1914}
F.~Hausdorff.
\newblock {\em Grundzüge der Mengenlehre}.
\newblock Veit\&Comp, Leipzig, 1914.

\bibitem{Jaffe2002}
C.~Jaff\'e, S.~D. Ross, M.~W. Lo, J.~Marsden, D.~Farrelly, and T.~Uzer.
\newblock Statistical theory of asteroid escape rates.
\newblock {\em Phys. Rev. Lett.}, 89:011101, Jun 2002.

\bibitem{koon_heteroclinic_2000}
W.~S. Koon, M.~W. Lo, J.~E. Marsden, and S.~D. Ross.
\newblock Heteroclinic connections between periodic orbits and resonance
  transitions in celestial mechanics.
\newblock {\em Chaos: An Interdisciplinary Journal of Nonlinear Science},
  10(2):427--469, jun 2000.

\bibitem{Koon2011}
W.~S. Koon, M.~W. Lo, J.~E. Marsden, and S.~D. Ross.
\newblock {\em {Dynamical systems, the three-body problem and space mission
  design}}.
\newblock Marsden books, 2011.

\bibitem{Lyu2020}
W.~Lyu, S.~Naik, and S.~Wiggins.
\newblock Uposham: A python package for computing unstable periodic orbits in
  two-degree-of-freedom hamiltonian systems.
\newblock {\em Journal of Open Source Software}, 5(45):1684, 2020.

\bibitem{Mendoza2010}
C.~Mendoza and A.~M. Mancho.
\newblock Hidden geometry of ocean flows.
\newblock {\em Phys. Rev. Lett.}, 105:038501, Jul 2010.

\bibitem{Meyer2009}
K.~R. Meyer, G.~R. Hall, and D.~Offin.
\newblock {\em {Introduction to Hamiltonian Dynamical Systems and the N-Body
  Problem}}.
\newblock Springer, 2009.

\bibitem{naik_geometry_2017}
S.~Naik and S.~D. Ross.
\newblock Geometry of escaping dynamics in nonlinear ship motion.
\newblock {\em Communications in Nonlinear Science and Numerical Simulation},
  47:48--70, jun 2017.

\bibitem{naik_finding_2019b}
S.~Naik and S.~Wiggins.
\newblock Finding normally hyperbolic invariant manifolds in two and three
  degrees of freedom with {H}\'enon-{H}eiles-type potential.
\newblock {\em Phys. Rev. E}, 100:022204, Aug 2019.

\bibitem{Pollak1980}
E.~Pollak, M.~S. Child, and P.~Pechukas.
\newblock Classical transition state theory: {A} lower bound to the reaction
  probability.
\newblock {\em The Journal of Chemical Physics}, 72(3):1669--1678, 1980.

\bibitem{Waalkens2005Escape}
H.~Waalkens, A.~Burbanks, and S.~Wiggins.
\newblock {Escape from planetary neighbourhoods}.
\newblock {\em Monthly Notices of the Royal Astronomical Society},
  361(3):763--775, 08 2005.

\bibitem{Waalkens2005}
H.~Waalkens, A.~Burbanks, and S.~Wiggins.
\newblock A formula to compute the microcanonical volume of reactive initial
  conditions in transition state theory.
\newblock {\em Journal of Physics A: Mathematical and Theoretical},
  38(45):l759, 2005.

\bibitem{waalkens2004direct}
H.~Waalkens and S.~Wiggins.
\newblock {Direct construction of a dividing surface of minimal flux for
  multi-degree-of-freedom systems that cannot be recrossed}.
\newblock {\em J. Phys. A: Math. Gen.}, 37(35):L435, 2004.

\bibitem{Geomodels}
H.~Waalkens and S.~Wiggins.
\newblock Geometrical models of the phase space structures governing reaction
  dynamics.
\newblock {\em Regular and Chaotic Dynamics}, 15, 06 2009.

\bibitem{Wiggins1990}
S.~Wiggins.
\newblock On the geometry of transport in phase space i. transport in
  k-degree-of-freedom hamiltonian systems, $2 \le k < \infty$.
\newblock {\em Physica D: Nonlinear Phenomena}, 44(3):471 -- 501, 1990.

\bibitem{Wiggins1994}
S.~Wiggins.
\newblock {\em Normally hyperbolic invariant manifolds in dynamical systems}.
\newblock Springer-Verlag, 1994.

\bibitem{wiggins2003applied}
S.~Wiggins.
\newblock {\em Introduction to Applied Nonlinear Dynamical Systems and Chaos},
  volume~2.
\newblock Springer Science \& Business Media, 2003.

\bibitem{NHIMinchem}
S.~Wiggins.
\newblock The role of normally hyperbolic invariant manifolds (nhims) in the
  context of the phase space setting for chemical reaction dynamics.
\newblock {\em Regular and Chaotic Dynamics}, 21:621--638, 11 2016.

\bibitem{Wiggins2017book}
S.~Wiggins.
\newblock {\em Ordinary Differential Equations}.
\newblock 12 2017.

\bibitem{wiggins_impenetrable_2001}
S.~Wiggins, L.~Wiesenfeld, C.~Jaffé, and T.~Uzer.
\newblock Impenetrable {Barriers} in {Phase}-{Space}.
\newblock {\em Physical Review Letters}, 86(24):5478--5481, June 2001.

\end{thebibliography}
\bibliographystyle{abbrv}  

\end{document}